\documentclass[bibyear]{aa} 

\usepackage{lscape}
\usepackage{float}
\usepackage{graphicx}
\usepackage{txfonts}
\begin{document} 
\title{ArT\'eMiS imaging of the filamentary infrared dark clouds G1.75-0.08 and G11.36+0.80: Dust-based physical properties of the clouds and their clumps\thanks{This publication is based on data acquired with the Atacama Pathfinder EXperiment (APEX) under programme ID 108.21W4. APEX is a collaboration between the Max-Planck-Institut f{\"u}r Radioastronomie, the European Southern Observatory,
and the Onsala Space Observatory. This work is also based on observations carried out under project number 015-21 with the IRAM 30-metre telescope. IRAM is supported by INSU/CNRS (France), MPG (Germany) and IGN (Spain).}}

   \author{O.~Miettinen\inst{1}, M. Mattern\inst{2}, and Ph. Andr\'e\inst{2}}

   \institute{Academy of Finland, Hakaniemenranta 6, P.O. Box 131, FI-00531 Helsinki, Finland \\ \email{oskari.miettinen@aka.fi} \and Laboratoire d'Astrophysique (AIM), CEA, CNRS, Universit\'e Paris-Saclay, Universit\'e Paris Diderot, Sorbonne Paris Cit\'e, 91191 Gif-sur-Yvette, France}

   \date{Received ; accepted}

\authorrunning{Miettinen et al.}
\titlerunning{ArT\'eMiS imaging of the IRDCs G1.75-0.08 and G11.36+0.8}

\abstract{Filamentary infrared dark clouds (IRDCs) are a useful class of interstellar clouds for studying the cloud fragmentation mechanisms on different spatial scales. Determination of the physical properties of the substructures in IRDCs can also provide useful constraints on the initial conditions and early stages of star formation, including those of high-mass stars.}{We aim to determine the physical characteristics of two filamentary IRDCs, G1.75-0.08 and G11.36+0.80, and their clumps. We also attempt to understand how the IRDCs are fragmented into clumps.}{We imaged the target IRDCs at 350~$\mu$m and 450~$\mu$m using the bolometer called  Architectures de bolom\`{e}tres pour des T\'elescopes \`{a} grand champ de vue dans le domaine sub-Millim\'etrique au Sol (ArT\'eMiS). These data were used in conjunction with our previous 870~$\mu$m observations with the Large APEX BOlometer CAmera (LABOCA) and archival \textit{Spitzer} and \textit{Herschel} data. The LABOCA clump positions in G11.36+0.80 were also observed in the N$_2$H$^+(1-0)$ transition with the Institut de Radioastronomie Millim\'etrique (IRAM) 30-metre telescope.}{On the basis of their far-IR to submillimetre spectral energy distributions (SEDs), G1.75-0.08 was found to be composed of two cold ($\sim14.5$~K), massive (several $\sim10^3$~M$_{\sun}$) clumps that are projectively separated by $\sim3.7$~pc. Both clumps are 70~$\mu$m dark, but they do not appear to be bounded by self-gravity. The G1.75-0.08 filament was found to be subcritical by a factor of $\sim14$ with respect to its critical line mass, but the result is subject to uncertain gas velocity dispersion. The IRDC G11.36+0.80 was found to be moderately (by a factor of $\sim2$) supercritical and composed of four clumps that are detected at all wavelengths observed with the ground-based bolometers. The SED-based dust temperatures of the clumps are $\sim13-15$~K, and their masses are in the range $\sim 232-633$~M$_{\sun}$. All the clumps are gravitationally bound and they appear to be in somewhat different stages of evolution on the basis of their luminosity-to-mass ratio. The projected, average separation of the clumps is $\sim1$~pc. At least three clumps in our sample show hints of fragmentation into smaller objects in the ArT\'eMiS images.}{A configuration that is observed in G1.75-0.08, namely two clumps at the ends of the filament, could be the result of gravitational focussing acting along the cloud. The two clumps fulfil the mass-radius threshold for high-mass star formation, but if their single-dish-based high velocity dispersion is confirmed, their gravitational potential energy would be strongly overcome by the internal kinetic energy, and the clumps would have to be confined by external pressure to survive. Owing to the location of G1.75-0.08 near the Galactic centre ($\sim270$~pc), environmental effects such as a high level of turbulence, tidal forces, and shearing motions could affect the cloud dynamics. The observed clump separation in G11.36+0.80 can be understood in terms of a sausage instability, which conforms to the findings in some other IRDC filaments. The G11.36+0.80 clumps do not lie above the mass-radius threshold where high-mass star formation is expected to be possible, and hence lower-mass star formation seems more likely. The substructure observed in one of the clumps in G11.36+0.80 suggests that the IRDC has fragmented in a hierarchical fashion with a scale-dependent physical mechanism. This conforms to the filamentary paradigm for Galactic star formation.}


   \maketitle
%

\section{Introduction}

Dust continuum imaging in the (sub-)millimetre regime ($\lambda=350-1\,000$~$\mu$m) is a very useful observational approach for 
characterising the physical properties of interstellar molecular clouds and their substructures, where the formation of new stars can take place. For the studies of the so-called infrared dark clouds (IRDCs), dust continuum observations are of particular importance because although the clouds are typically selected as dark absorption features against the Galactic mid-IR background (\cite{perault1996}; 
\cite{egan1998}; \cite{simon2006}; \cite{peretto2009}), detection of dust emission from the cloud can distinguish it as a true, physical IRDC from a local mimimum in the Galactic mid-IR background that might only resemble an IRDC (\cite{wilcock2012}). 

Infrared dark clouds themselves are very useful target sources for the studies of molecular cloud fragmentation and star formation. One reason for this is that IRDCs are at an early stage of evolution (e.g. \cite{rathborne2006}), and hence their initial fragmentation processes can be studied before any strong feedback from the newly formed stars has affected the cloud. Second, IRDCs often exhibit filamentary morphology with substructures of elevated density along the long axis of the filament (e.g. \cite{jackson2010}; \cite{kainulainen2013}; \cite{henshaw2016b}; \cite{miettinen2018}). Among the different fragmentation mechanisms that can operate in interstellar clouds, the so-called sausage instability with a particular length scale has been able to successfully explain the observed fragment separations in many IRDC filaments (e.g. the Nessie nebula (\cite{jackson2010}), G14.225-0.506 (\cite{busquet2013}), the Snake IRDC (\cite{kainulainen2013}), G34.43+00.24 (\cite{foster2014}), and the Seahorse IRDC (\cite{miettinen2018})). Moreover, some filamentary IRDCs are found to exhibit a hierarchical fragmentation behaviour with spatial-scale-dependent fragmentation mechanism (e.g. \cite{kainulainen2013}; \cite{wang2014}; \cite{mattern2018a}; \cite{liu2021}). Similarly, other Galactic filamentary molecular clouds such as those in Orion (e.g. \cite{takahashi2013}; \cite{teixeira2016}) and NGC~6334 (e.g. \cite{shimajiri2019}) are found to be fragmented in a hierarchical fashion, which suggests that the fragmentation of the filament proceeds at least in a bimodal way (from filaments to clumps, and from clumps to cores), and that this is a qualitatively similar or quasi-similar characteristic over a wide range of filament masses and densities (see \cite{andre2014} for a review; \cite{andre2019}).

A third reason for IRDCs being useful target sources is that some of the IRDC population members show evidence of ongoing high-mass ($M>8$~M$_{\sun}$) star formation (e.g. \cite{rathborne2006}; \cite{beuther2007}; \cite{chambers2009}; \cite{battersby2010}). In this process, many details and even the exact mechanism(s) are still to be deciphered (e.g. \cite{motte2018} for a review).

In this paper, we present the results of our dual-band (350~$\mu$m and 450~$\mu$m) submillimetre dust continuum imaging of two filamentary IRDCs, namely G1.75-0.08 and G11.36+0.80. Selected positions in G11.36+0.80 were also observed in the $J=1-0$ transition of N$_2$H$^+$, which provides us with information about the gas kinematics in the cloud. The target IRDCs were already imaged with the Large APEX BOlometer CAmera (LABOCA; \cite{siringo2009}) at 870~$\mu$m by Miettinen (2012), and they were part of the sample in the molecular line study of IRDCs by Miettinen (2014) that was based on the Millimetre Astronomy Legacy Team 90~GHz (MALT90) survey (\cite{foster2011}, 2013; \cite{jackson2013}). The angular resolution of the new dust continuum data is about twice better than that of our previous LABOCA data, and the new wavelength bands allow us to construct 
the spectral energy distributions (SEDs) of the detected sources and hence determine their physical properties more accurately than previously. 

The observations, data reduction, and ancillary data are described in Sect.~2. The analysis and results are described in Sect.~3. We discuss the results in Sect.~4, while in Sect.~5, we summarise our results and main conclusions. 

\section{Observations, data reduction, and ancillary data}

\subsection{ArT\'eMiS submillimetre continuum imaging}

Target fields with sizes of $8\farcm13 \times 5\farcm05$ and $6\farcm37 \times 7\farcm08$ and centred on the IRDCs G1.75-0.08 and G11.36+0.80, respectively, were imaged with the bolometer called Architectures de bolom\`{e}tres pour des T\'elescopes \`{a} grand champ de vue dans le domaine 
sub-Millim\'etrique au Sol (ArT\'eMiS; see \cite{reveret2014}; \cite{andre2016}; \cite{talvard2018}) on the Atacama Pathfinder EXperiment (APEX; \cite{gusten2006}) telescope. The ArT\'eMiS bolometer array consists of 2\,300 individual bolometers, and it currently operates simultaneously at two wavebands, namely 350~$\mu$m and 450~$\mu$m. The angular resolutions (half-power beam width, or HPBW) at these wavelengths are $8\farcs5$ and $9\farcs4$, respectively. The field of view (FoV) of the array is $4\farcm7 \times 2\farcm5$ at both wavelengths.

Our observations were carried out on 21--23 August 2021 under very good weather conditions with a typical precipitable water vapour (PWV) of 0.4~mm. The zenith opacity at the ArT\'eMiS frequencies was measured with skydip observations and was found to lie in the range $\tau_{\rm z}=0.26-0.78$. The observations were performed using the total power on-the-fly (OTF) mapping mode with the dump and grid spacings (i.e. along and perpendicular to the scanning direction) set to $8\arcsec$ and $4\arcsec$, respectively. The integration time per dump was one second. We note that ArT\'eMiS uses filled bolometer arrays that essentially fully sample the aforementioned FoV at any given time. Scanning was performed in only one direction, namely along the right ascension axis. This can lead to the beam smearing effects and artefacts in the maps (see Sect.~4.5 for further discussion). Owing to the need of subtracting correlated sky noise, the spatial filtering scale for the ArT\'eMiS data is typically $\sim2\arcmin$, which is comparable to the smaller diameter of the FoV (\cite{andre2016}).

The telescope focus and pointing were optimised and checked at regular intervals on the moon Ganymede, the planets Uranus and Neptune, and multiple different massive star-forming objects (G5.89-0.39, G10.47+0.03, G10.62-0.38, G34.26+0.15, and G45.07+0.13). The absolute flux calibration uncertainty is estimated to be $\sim30\%$ at 350~$\mu$m and $\sim20\%$ at 450~$\mu$m. Five hours of telescope time (including overheads) were spent on G1.75-0.08, while 5.8~hr was used for observations towards G11.36+0.80. Hence, the total telescope time used for the project was 10.8~hr. 

\begin{figure*}[!htb]
\begin{center}
\includegraphics[width=\textwidth]{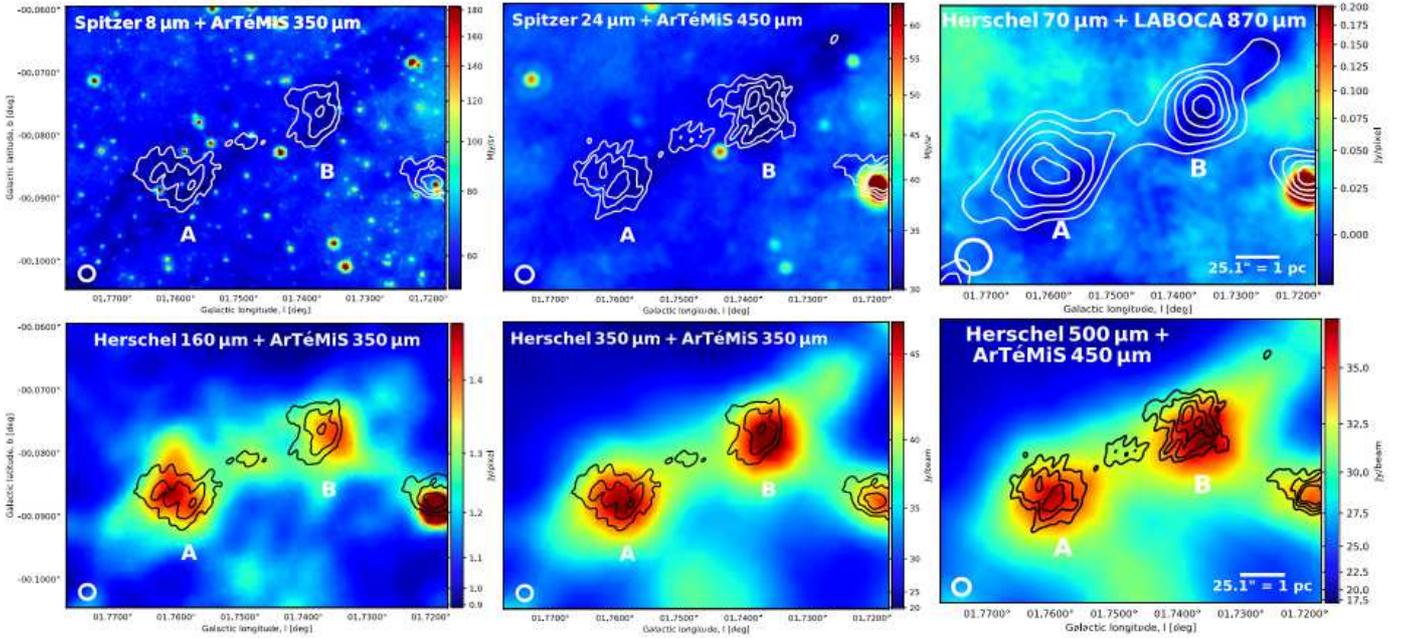}
\caption{Multi-wavelength images towards G1.75-0.08. In the top panels, the colour images from left to right show the \textit{Spitzer}/IRAC 8~$\mu$m, \textit{Spitzer}/MIPS 24~$\mu$m, and \textit{Herschel}/PACS 70~$\mu$m maps. The overlaid contours show the ArT\'eMiS 350~$\mu$m, ArT\'eMiS 450~$\mu$m, and LABOCA 870~$\mu$m emission, respectively. In each case, the contour levels start from $3\sigma$ and progress in steps of $1\sigma$. In the bottom panels, the colour images from left to right show the \textit{Herschel}/PACS 160~$\mu$m, \textit{Herschel}/SPIRE 350~$\mu$m, and \textit{Herschel}/SPIRE 500~$\mu$m maps. The first two are overlaid with the ArT\'eMiS 350~$\mu$m contours, while the rightmost map is overlaid with the ArT\'eMiS 450~$\mu$m contours. The beam sizes (HPBW) of the APEX bolometer observations are shown in the bottom left corner of each panel. A scale bar of 1~pc is shown in the right panels.}
\label{figure:G175images}
\end{center}
\end{figure*}

\begin{figure*}[!htb]
\begin{center}
\includegraphics[width=\textwidth]{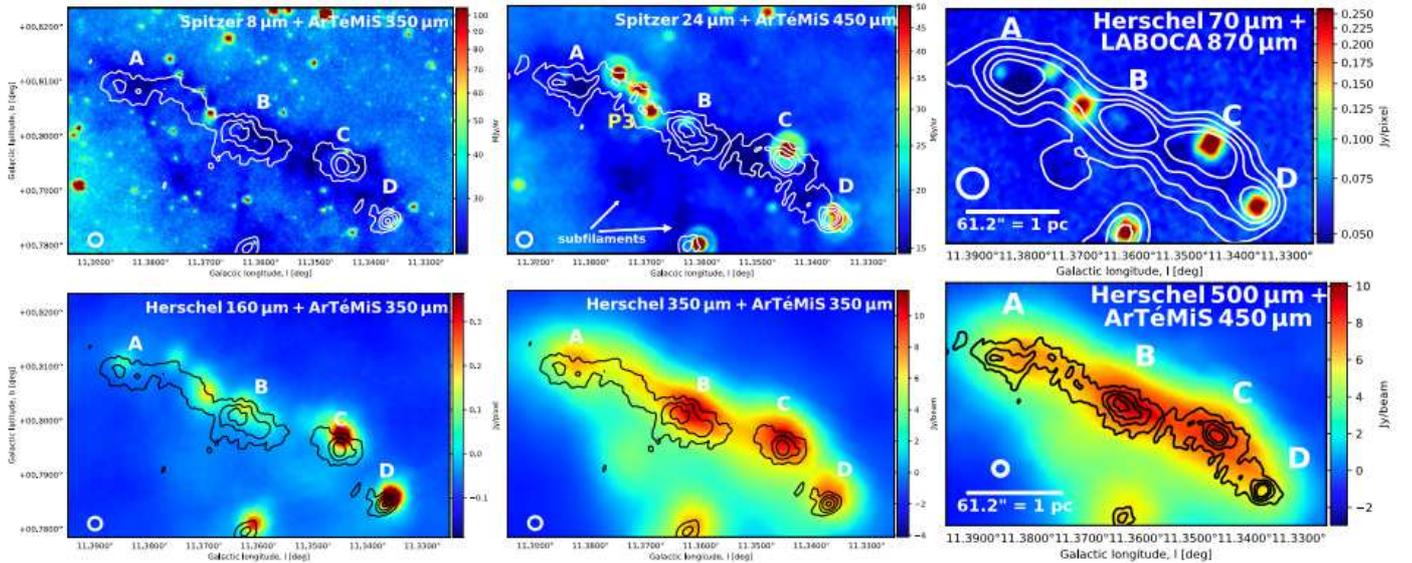}
\caption{Same as Fig.~\ref{figure:G175images}, but towards G11.36+0.80. All the contour levels start from $3\sigma$ and progress in steps of $3\sigma$. A source from the Feng et al. (2020) sample, G11.38+0.81 P3, is labelled P3 in the \textit{Spitzer} 24~$\mu$m plus ArT\'eMiS 450~$\mu$m image. The 24~$\mu$m dark features to the south of clump B are also indicated in this image. We call them subfilaments.}
\label{figure:G1136images}
\end{center}
\end{figure*}

The data were reduced using a dedicated data reduction package in \texttt{IDL} (Interactive Data Language). The \texttt{IDL} packages are publicly available on the APEX website\footnote{\url{www.apex-telescope.org/ns/data-reduction-for-bolometer-arrays/#ArtemisPipeline}.}. 
The $1\sigma$ rms noise levels in the final 350~$\mu$m and 450~$\mu$m maps of G1.75-0.08 were determined to be 0.88~Jy~beam$^{-1}$ and 0.53~Jy~beam$^{-1}$, respectively, while those for G11.36+0.80 are 0.43~Jy~beam$^{-1}$ and 0.29~Jy~beam$^{-1}$. The ArT\'eMiS contour maps of the target IRDCs are shown in Figs.~\ref{figure:G175images} and \ref{figure:G1136images} along with multi-wavelength images of the IRDCs.

\subsection{IRAM 30-metre telescope observations}

The LABOCA 870~$\mu$m peak positions along the G11.36+0.80 filament, which we call clumps~A--D (Fig.~\ref{figure:G1136images}), were observed in the $J=1-0$ transition of N$_2$H$^+$ with the Institut de Radioastronomie Millim\'etrique (IRAM) 30-metre telescope. The observations were carried out on 17 October 2021 during a pool observing week. 

As a front end, we used the Eight MIxer Receiver (EMIR; \cite{carter2012}) band E090. The backend was the Versatile SPectrometer Array (VESPA), where the bandwidth was $18\times20$~MHz with a corresponding channel spacing of 20~kHz. The N$_2$H$^+(1-0)$ line was tuned at 93\,173.7637~MHz, which corresponds to the frequency of the strongest hyperfine component ($J_{F_1,\,F}=1_{2,\,3}-0_{1,\,2}$) of the transition (\cite{pagani2009}). The aforementioned channel spacing yielded a velocity resolution of about 64~m~s$^{-1}$. The telescope beam size (HPBW) at the observed frequency is $26\farcs4$.

The observations were performed in the frequency-switching mode with a frequency throw of $\pm3.9$~MHz. Clumps A and B were observed for 22.5~min each, while clumps C and D were observed for 28.1~min. The telescope focus and pointing
were optimised and checked on the planet Venus and G5.89-0.39. The PWV during the observations was measured to be between
6.8~mm and 8.6~mm. The system temperatures during the observations were within the range of $T_{\rm sys} = 103-108$~K. 
The observed intensities were converted into the main-beam brightness temperature scale by using a main-beam efficiency factor of 
$\eta_{\rm MB}=0.80$.

The spectra were reduced using the Continuum and Line Analysis Single-dish Software ({\tt CLASS90}) of the GILDAS
software package (version mar19b)\footnote{Grenoble Image and Line Data Analysis Software (GILDAS)
is provided and actively developed by IRAM, and is available at \url{www.iram.fr/IRAMFR/GILDAS/}.}. 
The individual spectra were averaged with weights $w_i \propto t_{\rm int}/T_{\rm sys}^2$, where $t_{\rm int}$ is the on-source integration time.
Linear baselines were determined from the velocity ranges free of spectral line features, and then subtracted from the spectra. The resulting $1\sigma$ rms noise levels were about 23--32~mK.

\begin{figure}[!htb]
\centering
\resizebox{\hsize}{!}{\includegraphics{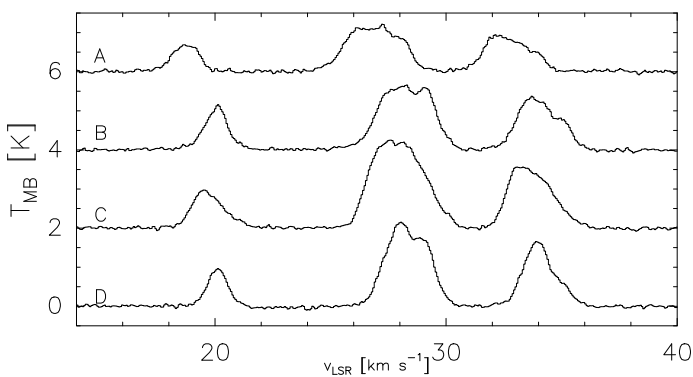}}
\caption{N$_2$H$^+(1-0)$ spectra towards the LABOCA clumps in G11.36+0.80. The intensity scale is shown in units of the main-beam brightness temperature.}
\label{figure:spectra}
\end{figure}

The reduced spectra are shown in Fig.~\ref{figure:spectra}. The $J=1-0$ transition of N$_2$H$^+$ is split into 15 hyperfine components, where the aforementioned strongest component has a relative intensity of $R_i = 7/15$. The hyperfine components are blended into three groups in the observed spectra. We fitted the hyperfine structure in {\tt CLASS90} using the rest frequencies from Pagani et al. (2009, Table~2 therein) and the relative intensities from Mangum \& Shirley (2015, Table~7 therein). The derived spectral line parameters are given in Table~\ref{table:line}. The listed parameters are the local standard of rest (LSR) radial velocity ($v_{\rm LSR}$), full width at half maximum (FWHM; $\Delta v$), peak intensity ($T_{\rm MB}$), and the total optical thickness of the line ($\tau$). 

\begin{table}[H]
\renewcommand{\footnoterule}{}
\caption{Parameters of the N$_2$H$^+(1-0)$ spectral lines detected towards the LABOCA clumps in G11.36+0.80.}
\begin{minipage}{1\columnwidth}
\centering
\label{table:line}
\begin{tabular}{c c c c c}
\hline\hline 
Source & $v_{\rm LSR}$ & $\Delta v$ & $T_{\rm MB}$ & $\tau$\\
       & [km s$^{-1}$] & [km s$^{-1}$] & [K] &  \\ 
\hline 
Clump A & $26.7\pm0.006$ & $1.46\pm0.02$ & $1.22\pm0.09$ & $6.76\pm0.27$ \\
Clump B & $28.1\pm0.003$ & $1.28\pm0.01$ & $1.69\pm0.07$ & $7.35\pm0.18$ \\
Clump C & $27.7\pm0.002$ & $1.71\pm0.08$ & $2.31\pm0.08$ & $2.29\pm0.06$ \\
Clump D & $28.1\pm0.002$ & $1.12\pm0.02$ & $2.12\pm0.04$ & $2.37\pm0.02$ \\	
\hline
\end{tabular} 
\end{minipage} 
\end{table}

\subsection{Ancillary data}

We complemented our new ArT\'eMiS submillimetre continuum data with the earlier LABOCA 870~$\mu$m imaging data of the target IRDCs from Miettinen (2012). We note that G1.75-0.08 was part of a larger target field of Miettinen (2012) called G1.87-0.14. In brief, the angular resolution of the LABOCA data is $19\farcs86$ (HPBW), and the $1\sigma$ rms noise of the G1.75-0.08 map is 90~mJy~beam$^{-1}$ and that for G11.36+0.80 is about 31~mJy~beam$^{-1}$ around the IRDC in the centre of the map. 

We also employed data from the \textit{Herschel Space Observatory} (\cite{pilbratt2010})\footnote{\textit{Herschel} is an ESA space observatory with science instruments provided by European-led Principal Investigator consortia and with important participation from NASA.}, which were measured with the Photodetector Array Camera and Spectrometer (PACS; \cite{poglitsch2010}) in the far-IR regime (70--160~$\mu$m) with nominal angular resolutions of $\sim6\arcsec-12\arcsec$, and the Spectral and Photometric Imaging Receiver (SPIRE; \cite{griffin2010}) in the far-IR to submillimetre regime (250--500~$\mu$m) with nominal angular resolutions of $18\arcsec-35\arcsec$. In particular, we made use of the data products from the \textit{Herschel} infrared Galactic Plane Survey (Hi-GAL; \cite{molinari2016a}). The PACS beams of the Hi-GAL observations are elongated along the scan direction, and have sizes of $5\farcs8\times 12\farcs1$ at 70~$\mu$m and $11\farcs4\times 13\farcs4$ at 160~$\mu$m.

Our study also makes use of the mid-IR observations obtained with the \textit{Spitzer Space Telescope}'s (\cite{werner2004}) Infrared Array Camera (IRAC; \cite{fazio2004}) and the Multiband Imaging Photometer for \textit{Spitzer} (MIPS; \cite{rieke2004}). We used the 8~$\mu$m images taken as part of the Galactic Legacy Infrared Mid-Plane Survey Extraordinaire (GLIMPSE; \cite{churchwell2009}), and 24~$\mu$m images from the MIPS Galactic Plane Survey (MIPSGAL; \cite{carey2009}). Moreover, we employed the 24~$\mu$m Point Source Catalogue of the Galactic Plane from \textit{Spitzer}/MIPSGAL (\cite{gutermuth2015}).

\section{Analysis and results}

\subsection{Submillimetre dust emission characteristics}

In Tables~\ref{table:G175sources} and \ref{table:G1136sources}, we list the properties of the ArT\'eMiS and LABOCA clumps in the 
IRDCs G1.75-0.08 and G11.36+0.80, respectively. The parameters given in the tables are the coordinates of the dust emission peak position, signal-to-noise ratio (S/N) of the peak position, peak surface brightness, flux density of the source integrated within the $3\sigma$ level of the emission, the effective radius, which is defined as $R_{\rm eff}=\sqrt{A_{\rm proj,\, 3\sigma}/\pi}$, where $A_{\rm proj,\, 3\sigma}$ is the projected area within the $3\sigma$ extension of the source, and the FWHM size of the source, which was measured over the largest diameter of the half-maximum contour. The quoted uncertainties of the surface brightness and flux density take the absolute calibration uncertainties into account. The uncertainties in the effective radii and FWHM sizes were propagated from the average of the $\pm$ uncertainties in the cloud's kinematic distance (Table~\ref{table:distances}, column~(2)). As seen in Table~\ref{table:G1136sources}, the non-deconvolved FWHM extension of clump~D in G11.36+0.80 is comparable to the beam size (HPBW) at each wavelength, and the corresponding deconvolved sizes make the source a point source. We note that the aforementioned method for determining the source flux density, that is, above the $3\sigma$ level, resembles the bijection paradigm presented by Rosolowsky et al. (2008) and the way how for example the {\tt clumpfind} algorithm (\cite{williams1994}) works.

We also determined the integrated flux densities of the whole filamentary structures of G1.75-0.08 and G11.36+0.80. The ArT\'eMiS and LABOCA flux densities that were all determined inside a region that corresponds to the LABOCA $3\sigma$ contour are listed in Table~\ref{table:irdcfluxes}. 

The \textit{Spitzer} and \textit{Herschel} counterparts of the clumps in G1.75-0.08 and G11.36+0.80 were searched from the NASA/IPAC Infrared Science Archive (IRSA)\footnote{\url{irsa.ipac.caltech.edu}.}. More details and the corresponding flux densities are given in Appendix~A.

\begin{table*}
\caption{Submillimetre dust continuum sources in G1.75-0.08.}
\begin{minipage}{2\columnwidth}
\centering
\renewcommand{\footnoterule}{}
\label{table:G175sources}
\begin{tabular}{c c c c c c c c c c}
\hline\hline 
Source & $\alpha_{2000.0}$ & $\delta_{2000.0}$ & S/N & $I$ & $S$ & \multicolumn{2}{c}{$R_{\rm eff}$} & \multicolumn{2}{c}{${\rm FWHM}_{\rm maj}$}\\
       & [h:m:s] & [$\degr$:$\arcmin$:$\arcsec$] & & [Jy~beam$^{-1}$] & [Jy] & [\arcsec] & [pc] & [\arcsec] & [pc]\\
\hline
\multicolumn{10}{c}{ArT\'eMiS 350 $\mu$m} \\
\hline
Clump A & 17 50 05.19 & -27 28 38.36 & 5.4 & $4.74\pm1.67$ & $42.6\pm13.2$ & 18.3 & $0.73\pm0.02$ & 45.5 & $1.81\pm0.04$\\
Clump B & 17 49 59.63 & -27 29 19.54 & 5.5 & $4.85\pm1.70$ & $33.3\pm10.6$ & 16.0 & $0.64\pm0.01$ & 39.3 & $1.57\pm0.03$\\
\hline
\multicolumn{10}{c}{ArT\'eMiS 450 $\mu$m} \\
\hline
Clump A & 17 50 05.78 & -27 28 30.75 & 6.0 & $3.16\pm0.63$ & $25.3\pm5.5$ & 19.3 & $0.77\pm0.02$ & 40.9 & $1.63\pm0.03$\\
Clump B & 17 49 59.50 & -27 29 20.75 & 6.8 & $3.60\pm0.72$ & $29.7\pm6.4$ & 20.4 & $0.81\pm0.02$ & 36.9 & $1.47\pm0.03$\\
\hline
\multicolumn{10}{c}{LABOCA 870 $\mu$m} \\
\hline
Clump A & 17 50 05.08 & -27 28 32.12 & 8.4 & $0.76\pm0.63$ & $4.05\pm0.60$ & 36.5 & $1.46\pm0.03$ & 49.4 & $1.97\pm0.04$\\
Clump B & 17 49 59.38 & -27 29 24.98 & 8.2 & $0.74\pm0.63$ & $2.73\pm0.50$ & 31.0 & $1.24\pm0.03$ & 42.6 & $1.70\pm0.04$\\
\hline 
\end{tabular} 
\tablefoot{The columns are as follows: (1) source name; (2) and (3) coordinates of the peak surface brightness position; (4) S/N ratio of the source; (5) peak surface brightness; (6) integrated flux density within the $3\sigma$ extension of the source; (7) and (8) effective radius of the source in angular and spatial units; (9) and (10) FWHM size of the source in angular and spatial units. The FWHM size refers to the largest diameter of the half-maximum contour, and the tabulated values are not deconvolved from the beam size.}
\end{minipage}
\end{table*}

\begin{table*}
\caption{Submillimetre dust continuum sources in G11.36+0.80.}
\begin{minipage}{2\columnwidth}
\centering
\renewcommand{\footnoterule}{}
\label{table:G1136sources}
\begin{tabular}{c c c c c c c c c c}
\hline\hline 
Source & $\alpha_{2000.0}$ & $\delta_{2000.0}$ & S/N & $I$ & $S$ & \multicolumn{2}{c}{$R_{\rm eff}$} & \multicolumn{2}{c}{${\rm FWHM}_{\rm maj}$}\\
       & [h:m:s] & [$\degr$:$\arcmin$:$\arcsec$] & & [Jy~beam$^{-1}$] & [Jy] & [\arcsec] & [pc] & [\arcsec] & [pc]\\
\hline
\multicolumn{10}{c}{ArT\'eMiS 350 $\mu$m} \\
\hline
Clump A & 18 07 37.22 & -18 41 09.92 & 7.4 & $3.20\pm1.03$ & $13.60\pm4.53$ & 13.6 & $0.21\pm0.02$ & 31.0 & $0.49\pm0.04$\\
Clump B & 18 07 36.30 & -18 42 29.91 & 11.8 & $5.06\pm2.00$ & $29.45\pm9.24$ & 17.4 & $0.27\pm0.02$ & 33.6 & $0.53\pm0.04$\\
Clump C & 18 07 35.39 & -18 43 41.53 & 11.8 & $5.09\pm2.01$ & $17.46\pm5.80$ & 14.1 & $0.22\pm0.02$ & 18.1 & $0.28\pm0.02$\\
Clump D & 18 07 36.68 & -18 44 24.96 & 13.2 & $5.67\pm2.38$ & $8.15\pm3.73$ & 9.16 & $0.14\pm0.01$ & 11.9 & $0.19\pm0.02$\\
\hline
\multicolumn{10}{c}{ArT\'eMiS 450 $\mu$m} \\
\hline
Clump A & 18 07 37.18 & -18 41 05.42 & 10.0 & $2.89\pm0.65$ & $17.11\pm3.73$ & 20.0 & $0.31\pm0.03$ & 32.6 & $0.51\pm0.04$\\
Clump B & 18 07 36.20 & -18 42 36.39 & 13.6 & $3.94\pm0.84$ & $20.27\pm4.52$ & 18.9 & $0.30\pm0.03$ & 30.3 & $0.47\pm0.04$\\
Clump C & 18 07 35.30 & -18 43 42.22 & 15.0 & $4.35\pm0.92$ & $20.95\pm4.50$ & 21.6 & $0.34\pm0.03$ & 17.1 & $0.27\pm0.02$\\
Clump D & 18 07 36.53 & -18 44 26.20 & 17.0 & $4.92\pm1.03$ & $10.77\pm2.87$ & 14.5 & $0.23\pm0.02$ & 13.5 & $0.21\pm0.02$\\
\hline
\multicolumn{10}{c}{LABOCA 870 $\mu$m} \\
\hline
Clump A & 18 07 36.69 & -18 41 13.51 & 19.9 & $0.62\pm0.07$ & $2.39\pm0.41$ & 37.8 & $0.59\pm0.05$ & 51.8 & $0.81\pm0.07$\\
Clump B & 18 07 35.86 & -18 42 41.75 & 24.0 & $0.75\pm0.08$ & $2.07\pm0.39$ & 31.9 & $0.50\pm0.04$ & 41.2 & $0.65\pm0.05$\\
Clump C & 18 07 35.04 & -18 43 45.75 & 19.9 & $0.62\pm0.07$ & $2.58\pm0.42$ & 35.3 & $0.55\pm0.05$ & 49.1 & $0.77\pm0.07$\\
Clump D & 18 07 36.14 & -18 44 25.74 & 19.2 & $0.60\pm0.07$ & $1.20\pm0.30$ & 26.2 & $0.41\pm0.03$ & 25.8 & $0.40\pm0.03$\\
\hline 
\end{tabular} 
\tablefoot{Columns are as in Table~\ref{table:G175sources}.}
\end{minipage}
\end{table*}

\subsection{Kinematic distances of the target IRDCs}

To determine the kinematic distances of the target IRDCs, we used the Monte Carlo technique of Wenger et al. (2018). This technique makes use of the 
Reid et al. (2014) rotation curve with updated solar motion parameters (see Table~2 in \cite{wenger2018})\footnote{The online calculation tool is available at \url{www.treywenger.com/kd/}.}. 

For G11.36+0.80, we used the mean value of the LSR velocities of N$_2$H$^+(1-0)$ listed in column~(2) in Table~\ref{table:line}, that is 
$27.65\pm0.33$~km~s$^{-1}$, where the uncertainty represents the standard error of the mean. For G1.75-0.08, we do not have single-pointing spectral line observations available, and hence we relied on the spectral line data presented in Miettinen (2014), which were taken as part of the 
MALT90 survey with the 22~m Mopra telescope in the OTF mapping mode. The HCN$(1-0)$ and HCO$^+(1-0)$ lines were the strongest detections of the spectra extracted towards clump~B in G1.75-0.08 (source G1.87-0.14 SMM~8 in Miettinen (2014); see Fig.~C.2 therein). Towards clump~A, HCN$(1-0)$ and HCO$^+(1-0)$ were also clearly detected, but the latter showed a clear double-peaked profile (source G1.87-0.14 SMM~15 in Miettinen (2014); see Fig.~C.6 therein). The mean value of the LSR velocities of HCN$(1-0)$ towards clump~A and HCN$(1-0)$ and HCO$^+(1-0)$ towards clump~B is $53.53\pm1.05$~km~s$^{-1}$, and this value was used in the kinematic distance calculation.

The derived distances are given in Table~\ref{table:distances}. The parameters listed in the table are the near and far kinematic distance solutions and the Galactocentric radius ($d_{\rm near}$, $d_{\rm far}$, and $R_{\rm GC}$, respectively). The quoted uncertainties represent the 68.3\% confidence interval. We assume that the IRDC lies at the near kinematic distance, where it is more likely to be seen in absorption against the Galactic mid-IR background radiation field. However, we note that G1.75-0.08 lies so close to the Galactic Centre ($\sim270$~pc) that the near and far distance solutions differ by only $\sim4\%$. 

\subsection{SED analysis}

The mid-IR or far-IR to submillimetre SEDs of the clumps in G1.75-0.08 and G11.36+0.80 were fitted by a modified blackbody (MBB) function using the method and assumptions of Miettinen (2020; Sect.~3.1 therein). For the 24~$\mu$m dark clumps, we used a single-temperature MBB model, while the SEDs of 24~$\mu$m bright clumps were fitted with a two-temperature MBB model. In brief, the SED analysis assumed a Ossenkopf \& Henning (1994) dust model for grains with thin ice mantles, where, for reference, the dust opacity at 870~$\mu$m is 1.38~cm$^2$~g$^{-1}$. The dust-to-total gas mass ratio was assumed to be 1/141 (i.e. 1.41 times lower than a dust-to-hydrogen mass ratio of 1/100, where the helium and metal mass contributions are taken to be 27\% and 2\%, respectively).

\begin{figure}[!htb]
\centering
\resizebox{.24\textwidth}{!}{\includegraphics{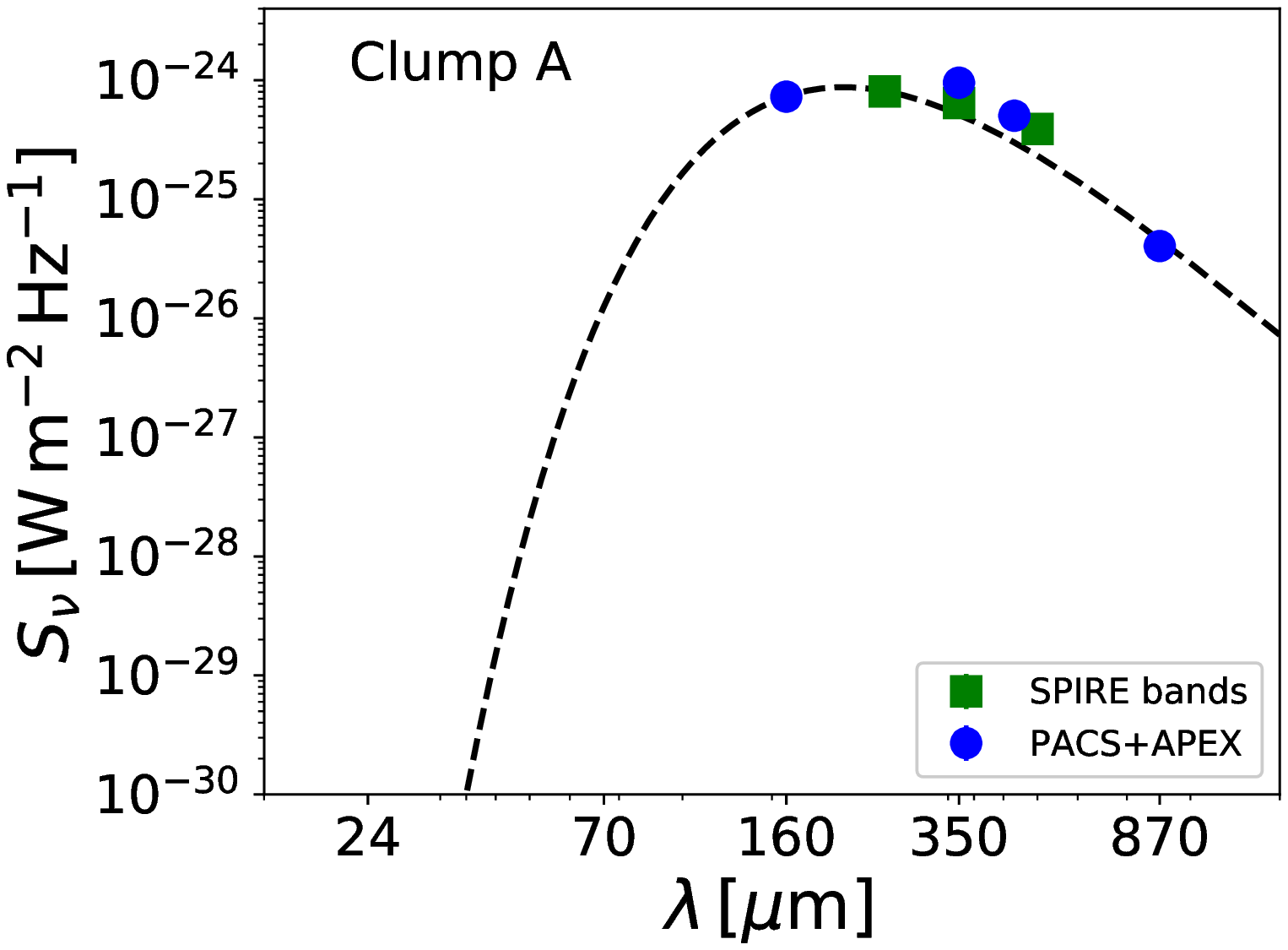}}
\resizebox{.24\textwidth}{!}{\includegraphics{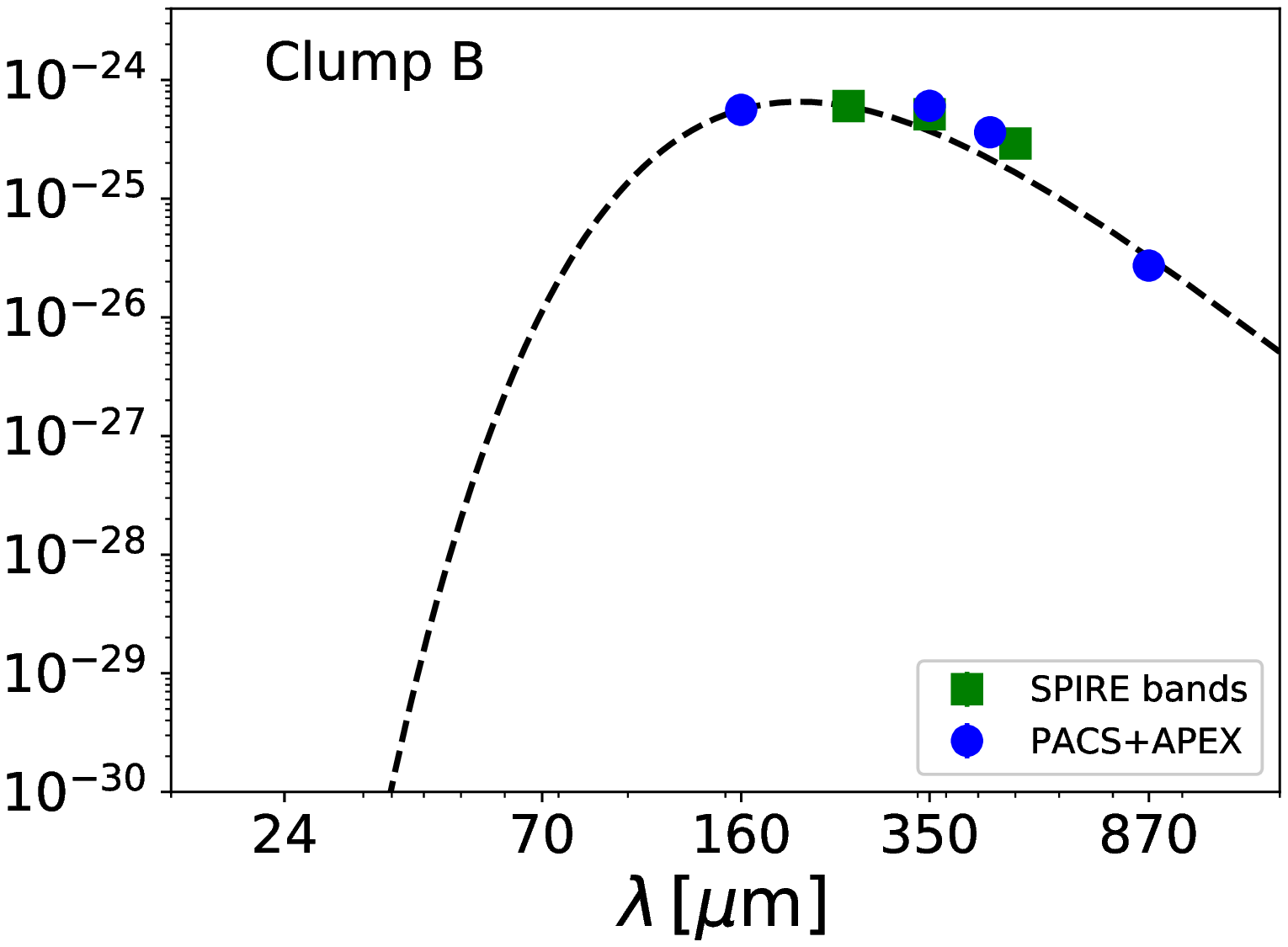}}
\caption{Far-IR to submillimetre SEDs of the clumps in G1.75-0.08. The SPIRE data points shown by square green symbols are plotted for comparison, and they were not included in the fit (see Sect.~3.3 for details). The dashed black lines represent the best MBB fits to the data.}
\label{figure:G175_seds}
\end{figure}

\begin{figure}[!htb]
\centering
\resizebox{.24\textwidth}{!}{\includegraphics{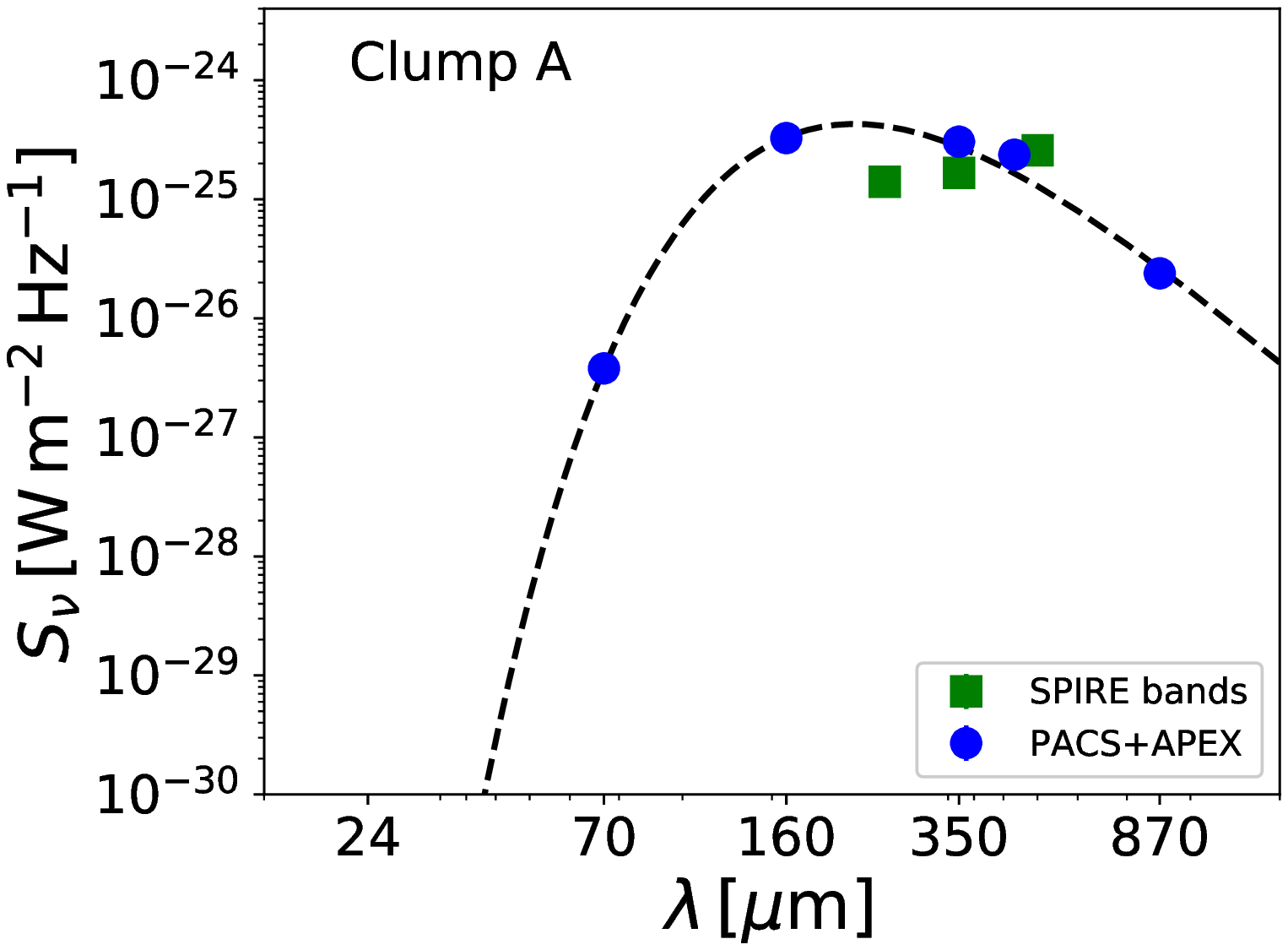}}
\resizebox{.24\textwidth}{!}{\includegraphics{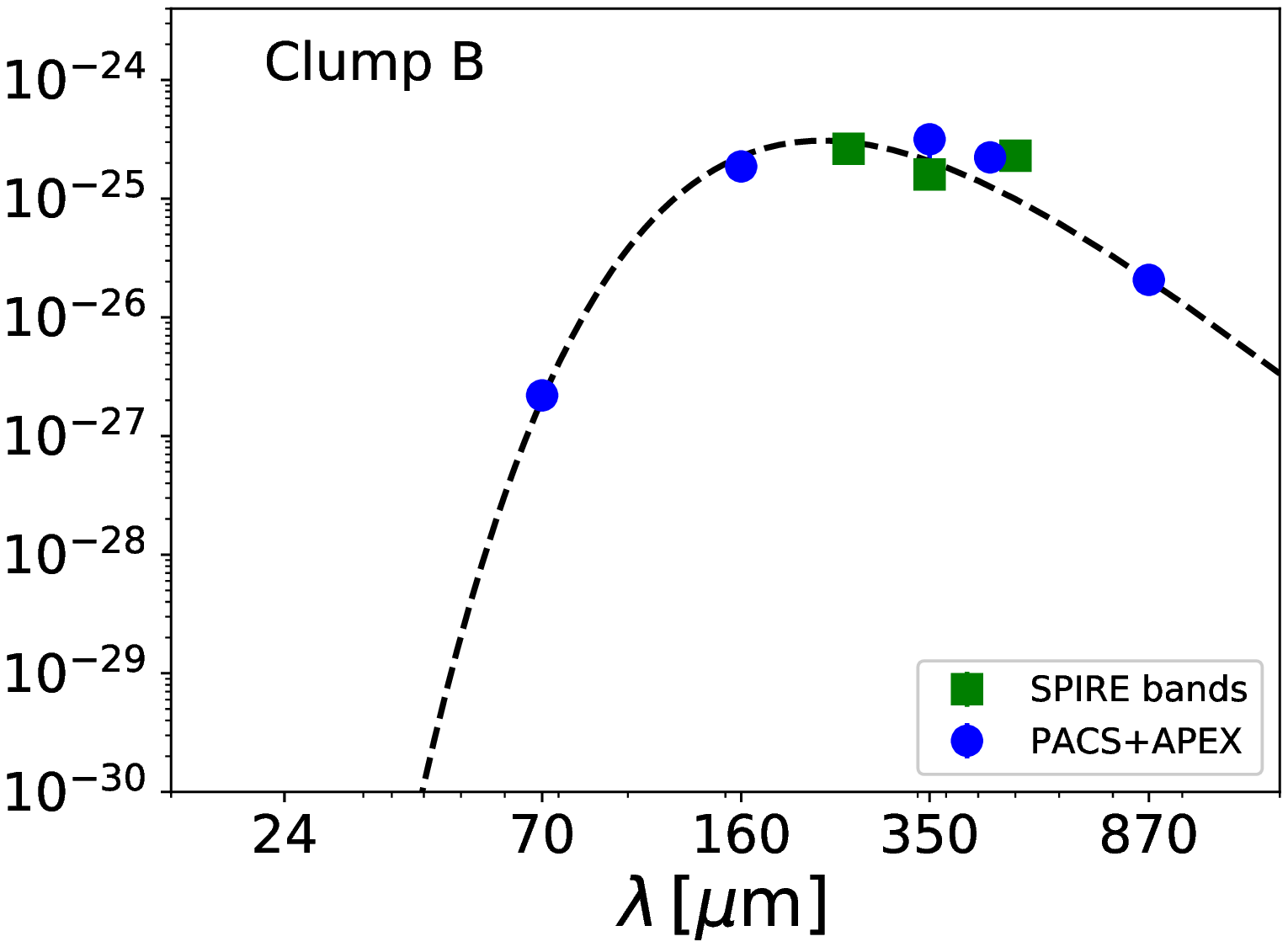}}
\resizebox{.24\textwidth}{!}{\includegraphics{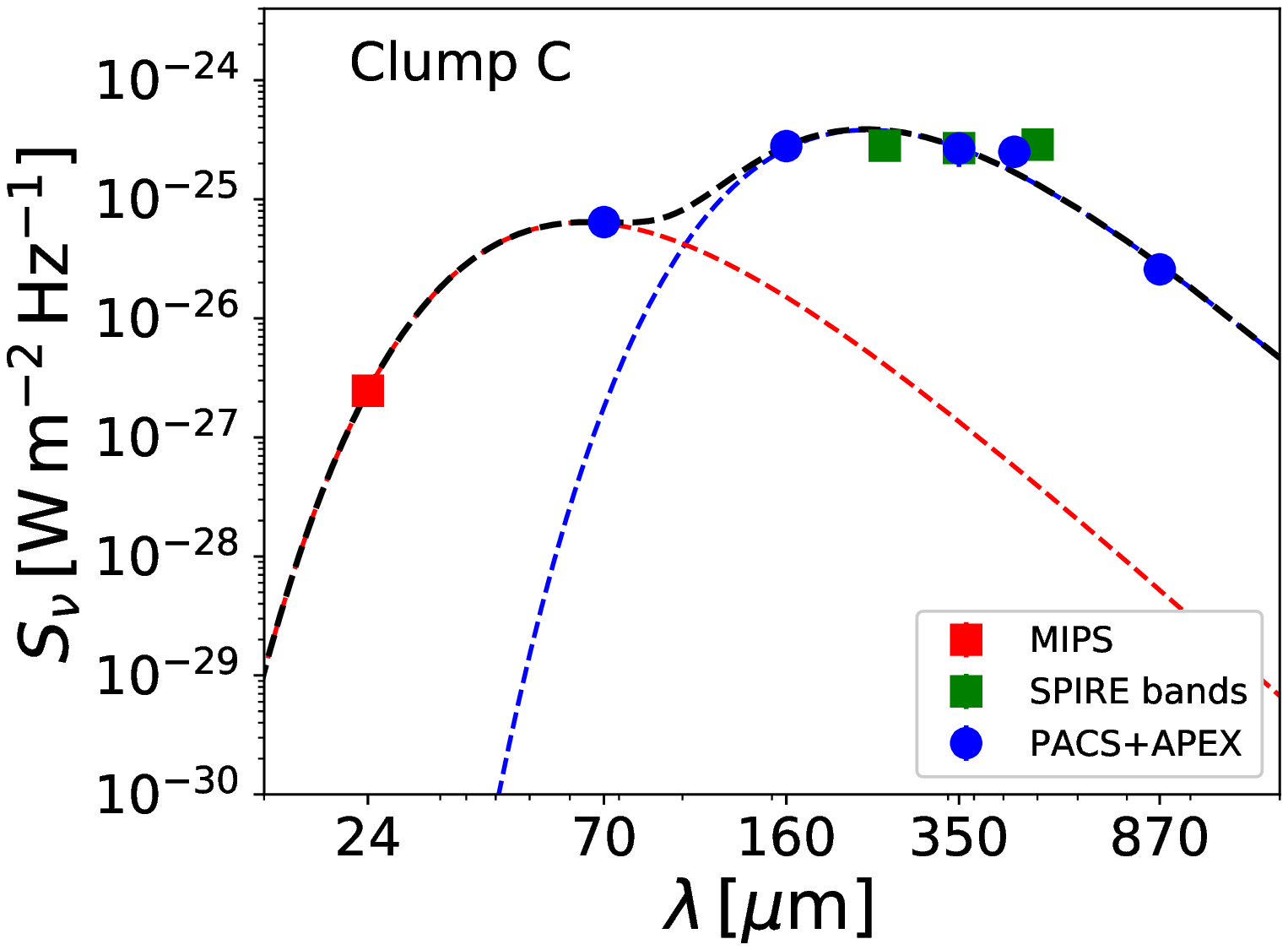}}
\resizebox{.24\textwidth}{!}{\includegraphics{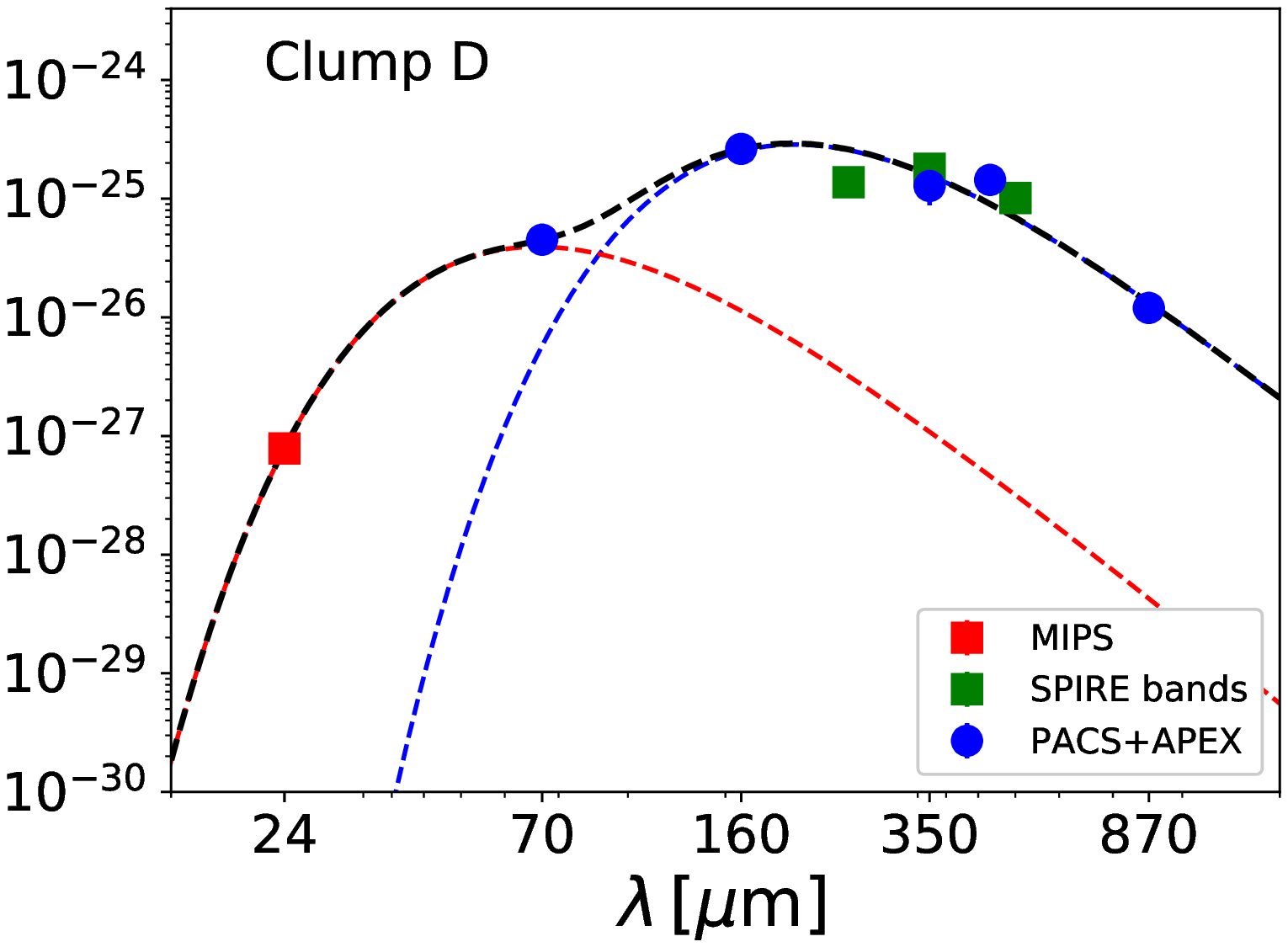}}
\caption{Mid-IR to submillimetre SEDs of the clumps in G11.36+0.80. The \textit{Spitzer}/MIPS 24~$\mu$m data points are highlighted in red. For the two-temperature MBB fits (clumps~C and D), the dashed black line shows the sum of the two components. The dashed blue and red lines show the SED fits to the cold and warm component, respectively.}
\label{figure:G1136_seds}
\end{figure}

\begin{figure}[!htb]
\centering
\resizebox{.24\textwidth}{!}{\includegraphics{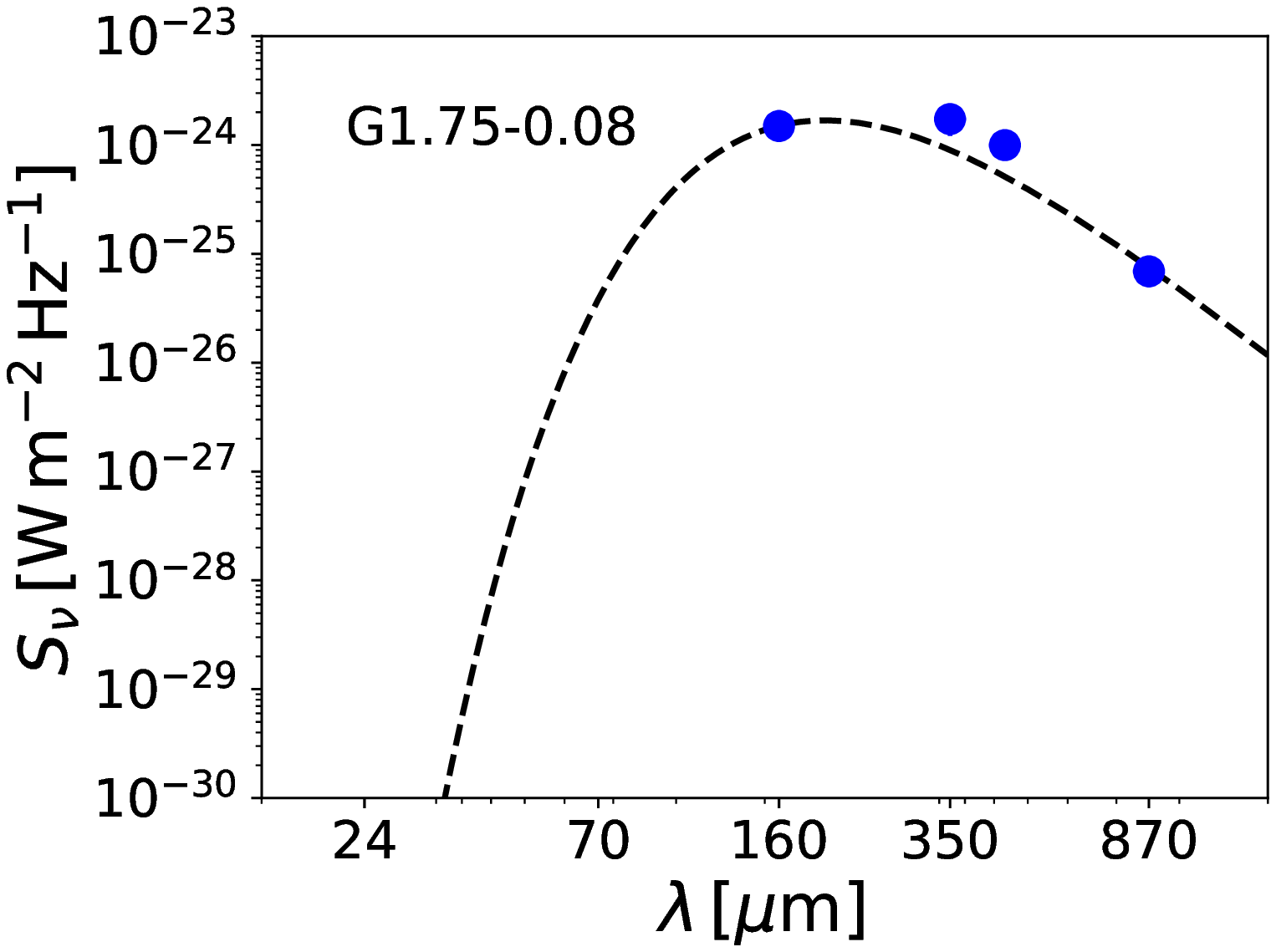}}
\resizebox{.24\textwidth}{!}{\includegraphics{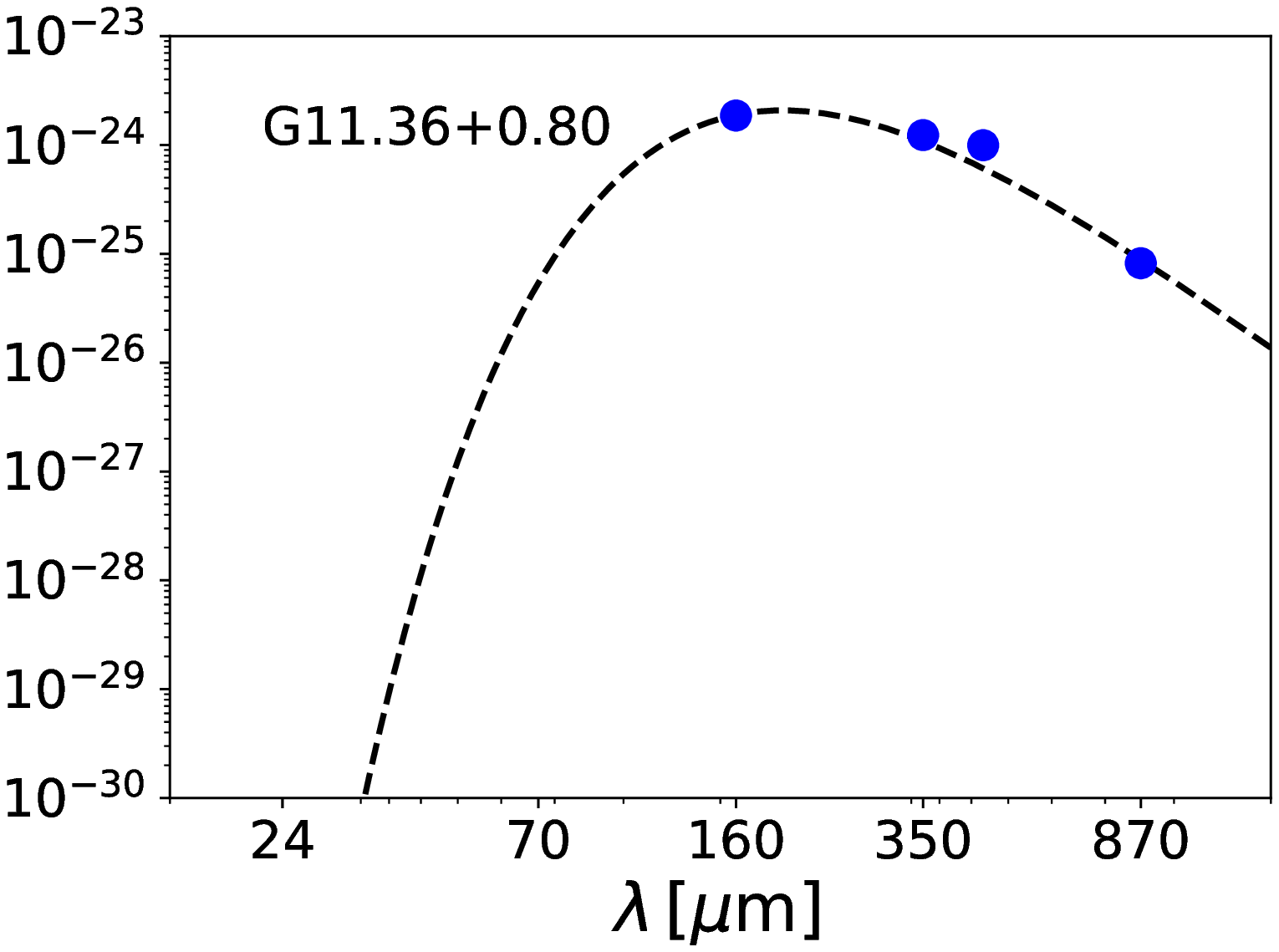}}
\caption{Far-IR to submillimetre SEDs of the G1.75-0.08 and G11.36+0.80 filaments.}
\label{figure:cloud_seds}
\end{figure}

The SEDs are shown in Figs.~\ref{figure:G175_seds} and \ref{figure:G1136_seds}. We note that the flux densities at $160-870$~$\mu$m were determined within apertures that corresponded to the LABOCA $3\sigma$ contour of emission so that the flux densities can be better compared to each other. However, the PACS 70~$\mu$m flux densities used in the fits were taken from the Hi-GAL catalogue (see Appendix~A) because those data are related to 70~$\mu$m point sources. The SPIRE $250-500$~$\mu$m data were not used in the fits because the 350~$\mu$m band is covered by our ArT\'eMiS data, the 500~$\mu$m band is close to our ArT\'eMiS 450~$\mu$m data point, and SPIRE is more sensitive to the extended dust emission than our ground-based bolometric data, and because the present work focusses on the high density parts of the filaments and their clumps.
Nevertheless, as discussed in Sect.~4.2, the SPIRE flux densities are mostly consistent with the SED fits. In Fig.~\ref{figure:cloud_seds} we show the SEDs of the G1.75-0.08 and G11.36+0.80 filaments that were constructed using the photometric data from Table~\ref{table:irdcfluxes} in conjunction with the PACS 160~$\mu$m flux densities integrated within the LABOCA $3\sigma$ contour. The 160~$\mu$m flux densities were determined as described in Appendix~A, that is, by determining the flux conversion factors on the basis of the Hi-GAL catalogue values, which also take the background emission contribution into account. We note that because our target IRDCs do not extend over a wide area in projection (the projected extensions are $\sim3\farcm4$ for G1.75-0.08 and $\sim4\farcm2$ for G11.36+0.80), the background (and foreground) emission is expected to be fairly uniform across the clouds.

The derived SED parameters are listed in Table~\ref{table:properties}. Columns~(2)--(7) in the table list the number of photometric data points used in the SED fit, reduced $\chi^2$ value of the fit, wavelength of the peak position of the fitted SED, dust temperature, total (gas+dust) mass of the source, and the bolometric luminosity integrated over the fitted SED curve. The quoted uncertainties were derived from the flux density uncertainties.  

\begin{table}[H]
\renewcommand{\footnoterule}{}
\caption{ArT\'eMiS and LABOCA flux densities of the G1.75-0.08 and G11.36+0.80 filaments.}
\begin{minipage}{1\columnwidth}
\centering
\label{table:irdcfluxes}
\begin{tabular}{c c c c c}
\hline\hline 
IRDC & $S_{\rm 350\, \mu m}$ & $S_{\rm 450\, \mu m}$ & $S_{\rm 870\, \mu m}$\\
       & [Jy] & [Jy] & [Jy] \\ 
\hline 
G1.75-0.08 & $172.9\pm51.9$ & $99.7\pm20.0$ & $6.9\pm0.8$ \\
G11.36+0.80 & $123.3\pm24.7$ & $99.5\pm19.9$ & $8.2\pm0.9$ \\
\hline
\end{tabular} 
\tablefoot{All the quoted flux densities were determined within the corresponding LABOCA $3\sigma$ contour level.}
\end{minipage} 
\end{table}

\begin{table}[H]
\renewcommand{\footnoterule}{}
\caption{Kinematic distances of the target IRDCs.}
\begin{minipage}{1\columnwidth}
\centering
\label{table:distances}
\begin{tabular}{c c c c}
\hline\hline 
IRDC & $d_{\rm near}$ & $d_{\rm far}$ & $R_{\rm GC}$\\
       & [kpc] & [kpc] & [kpc] \\ 
\hline 
G1.75-0.08 & $8.22^{+0.26}_{-0.09}$ & $8.55^{+0.12}_{-0.19}$ & $0.27^{+0.02}_{-0.02}$\\[1ex]
G11.36+0.80 & $3.23^{+0.34}_{-0.21}$ & $12.88^{+0.50}_{-0.25}$ & $5.12^{+0.28}_{-0.28}$ \\
\hline
\end{tabular} 
\tablefoot{The quoted uncertainties represent the 68.3\% confidence interval.}
\end{minipage} 
\end{table}

\begin{table*}
\caption{SED parameters and physical properties of the sources.}
\begin{minipage}{2\columnwidth}
\centering
\renewcommand{\footnoterule}{}
\label{table:properties}
\begin{tabular}{c c c c c c c c c c c}
\hline\hline
Source & $n_{\rm SED}^{\rm fit}$ & $\chi_{\rm red}^2$ & $\lambda_{\rm peak}$ & $T_{\rm dust}$ & $M$ & $L_{\rm bol}$ & $N({\rm H_2})$ & $n({\rm H_2})$ & $l$ & $R_{\rm fil}$\\
       & & & [$\mu$m] & [K] & [M$_{\sun}$] & [L$_{\sun}$] & [$10^{23}$ cm$^{-2}$] & [$10^3$ cm$^{-3}$] &  [pc] & [pc]\\
\hline 
\multicolumn{11}{c}{G1.75-0.08} \\
\hline
Filament & 4 & 4.49 & 200.0 & $15.0\pm0.4$ & $8\,289\pm1\,196$ & $6\,358^{+2\,189}_{-1\,749}$ & $0.3\pm0.05$ & $9.3\pm1.3$ &  $8.2\pm0.2$ & $0.70\pm0.01$\\[1ex]
Clump A & 4 & 3.55 & 214.3 & $14.3\pm0.4$ & $5\,501\pm935$ & $3\,140^{+1\,237}_{-965}$ & $1.7\pm0.6$ & $6.1\pm1.0$ &  \ldots & \ldots\\[1ex]
Clump B & 4 & 3.30 & 200.0 & $14.5\pm0.5$ & $3\,770\pm738$ & $2\,402^{+1\,107}_{-833}$ & $1.7\pm0.6$ & $6.8\pm1.3$ &  \ldots & \ldots\\
\hline 
\multicolumn{11}{c}{G11.36+0.80} \\
\hline
Filament & 4 & 2.31 & 200.0 & $15.2\pm0.5$ & $1\,468\pm199$ & $1\,232^{+432}_{-342}$ & $0.2\pm0.03$ & $11.5\pm1.6$ & $3.9\pm0.3$ & $0.39\pm0.03$ \\[1ex]
Clump A & 5 & 0.90 & 214.3 & $13.6\pm0.1$ & $532\pm56$ & $226^{+39}_{-34}$ & $1.3\pm0.4$ & $8.9\pm0.9$ & \ldots & \ldots\\[1ex]
Clump B & 5 & 2.60 & 230.8 & $13.2\pm0.2$ & $437\pm66$ & $157^{+41}_{-36}$ & $2.3\pm0.9$ & $12.0\pm1.8$ & \ldots & \ldots\\[1ex]
Clump C\tablefootmark{a} & 6 & 1.41 & 230.8 & $12.8\pm0.3$ & $633\pm112$ & $344^{+88}_{-182}$ & $2.6\pm1.0$ & $13.0\pm2.3$ & \ldots & \ldots\\[1ex]
        &  \ldots & \ldots & \ldots & $46.5\pm0.1$ & $0.19\pm0.01$ &   \ldots             &  \ldots & \ldots & \ldots & \ldots\\[1ex]
Clump D\tablefootmark{a} & 6 & 1.87 & 200.0 & $14.8\pm0.6$ & $232\pm56$ & $263^{+117}_{-87}$ & $1.8\pm0.8$ & $11.5\pm2.8$ & \ldots & \ldots\\
        &  \ldots & \ldots & \ldots & $43.1\pm0.3$ & $0.17\pm0.01$ &    \ldots            &  \ldots & \ldots & \ldots & \ldots \\[1ex]
\hline
\end{tabular} 
\tablefoot{The parameters given in the table are the number of flux density data points used in the SED fit, reduced $\chi^2$ value, wavelength of the peak position of the fitted SED, dust temperature, total (gas+dust) mass, bolometric luminosity, H$_2$ column density (source-averaged value within the LABOCA $3\sigma$ emission for the filaments, and for the clumps in their ArT\'eMiS 350~$\mu$m peak position), volume-averaged H$_2$ number density, projected length of the filament, and the outer radius of the filament (see Sect.~3.4).\tablefoottext{a}{The SED of the source was fitted with a two-temperature MBB model. The temperature and mass of the warm component are shown below the cold dust component values.}   }
\end{minipage} 
\end{table*}

\subsection{H$_2$ column and number densities}

The H$_2$ column densities of the clumps were calculated from the ArT\'eMiS 350~$\mu$m peak surfaces brightness and using the SED-based dust temperature of the clump (see e.g. Eq.~(3) in Miettinen et al. (2009)). The dust temperature of the cold component was adopted for clumps~C and D in G11.36+0.80. The mean molecular weight per H$_2$ molecule needed in the calculation was assumed to be 2.82, which results from the hydrogen mass percentage of 71\% assumed in the calculation of the dust-to-gas mass ratio (Sect.~3.3). The volume-averaged H$_2$ number densities of the clumps were calculated assuming a spherical geometry with the effective radius listed in column~(8) in Tables~\ref{table:G175sources} and \ref{table:G1136sources}. We note that the assumption of a uniformly distributed mass can be too simplistic for a source with a density distribution, and the central density can significantly exceed the volume-averaged value (see Sect.~4.3.1). 

We also calculated the H$_2$ column and number densities of the target IRDC filaments. The former was calculated by dividing the mass of the filament by the area enclosed by the LABOCA $3\sigma$ contour and using the aforementioned mean molecular weight value. To calculate the number densities of the filaments, we assumed a cylindrical geometry, where the length of the filament was taken to be the projected length of the long axis of the LABOCA emission at $3\sigma$ level. These are about 8.2~pc for G1.75-0.08 and 3.9~pc for G11.36+0.80. The outer radius of the filament was calculated as the effective radius by assuming a rectangular geometry where the area is equal to that enclosed by the LABOCA $3\sigma$ contour and the length the aforementioned extension of the LABOCA $3\sigma$ emission. This yielded outer radii of about 0.70~pc and 0.39~pc for G1.75-0.08 and G11.36+0.80, respectively. In this way, the filament size refers to the aperture sizes used to derive the flux densities and hence the mass of the filament. 

The derived H$_2$ column and number densities are listed in columns~(8) and (9) in Table~\ref{table:properties}. The quoted uncertainties in the $N({\rm H_2})$ values of the clumps were propagated from the 350~$\mu$m intensity and dust temperature uncertainties, while those of the filaments were propagated from the mass uncertainties. The uncertainties in $n({\rm H_2})$ were propagated from the mass uncertainties and for the clumps also from the uncertainties of the clump radius. The sizes (projected length and outer radius) of the filaments are listed in columns~(10) and (11) in Table~\ref{table:properties}.

\subsection{Virial parameter analysis}

The virial parameter of the clumps, which is defined as the ratio of the virial mass to the clump mass, was calculated using the same approach as in Miettinen (2020). The clumps were assumed to have a radial density profile of the form $n(r)\propto r^{-p}$ with $p=1.6$, which is consistent with those derived for Galactic high-mass star-forming clumps, although we note that within the usual range of the density power-law indices, $p=1.5-2$, a given virial mass would vary by only 34\% (see \cite{miettinen2020} and references therein). The density profiles of our target filaments and clumps could be derived through combining the available dust continuum data (including the combination of the ArT\'eMiS 350~$\mu$m and SPIRE 350~$\mu$m data to probe the different spatial scales), but an analysis like this is beyond the scope of this study.

To calculate the total (thermal plus non-thermal) one-dimensional velocity dispersion, we assumed that the gas kinetic temperature is equal to the 
dust temperature derived from the SED fit. The mean molecular weight per free particle was assumed to be 2.37 to be consistent with the assumptions in the SED analysis. For the clumps in G11.36+0.80, the spectral line width was taken to be the FWHM of N$_2$H$^+(1-0)$ detected in the present study (column~(3) in Table~\ref{table:line}). For the G1.75-0.08 clumps, we used the FWHM of HCN$(1-0)$ ($23.40\pm1.24$~km~s$^{-1}$ for clump~A and $13.50\pm0.38$~km~s$^{-1}$ for clump~B) from Miettinen (2014). The HCN$(1-0)$ transition was the strongest line detected towards clump~B, while towards clump~A, the strongest line was that of HNCO, which exhibited a very broad profile with an FWHM of $30.40\pm1.39$~km~s$^{-1}$, however, possibly as a result of shocks (e.g. \cite{yu2018}). Hence, HCN$(1-0)$ was selected as a probe of the gas kinematics for clump~A as well.

The derived virial masses and virial parameters are given in Table~\ref{table:virial}. The quoted virial mass uncertainties were propagated from the temperature and line width (FWHM) uncertainties, and those uncertainties along with the clump mass uncertainties were propagated to the virial parameter uncertainties. 


\section{Discussion}

\subsection{General features of the dust emission and the clumps detected with the APEX bolometers}

The IRDC G1.75-0.08 is composed of two dust clumps separated in projection by about 3.7~pc (as measured between the LABOCA 870~$\mu$m maxima). The clumps appear to be connected by a dust-emitting bridge where the central part is also detected in our ArT\'eMiS images. If we count the 450~$\mu$m fragment in between clumps~A and B as a third fragment, the average 450~$\mu$m clump separation would become $1.9\pm0.2$~pc. The dust clumps in G1.75-0.08 do not appear to have \textit{Spitzer} IR counterparts and the clumps also appear dark in the PACS 70~$\mu$m image, although some diffuse 70~$\mu$m emission can be seen towards clump~A. 

The IRDC G11.36+0.80 is filamentary in projected shape, and it has four prominent LABOCA 870~$\mu$m clumps along its long axis that are all detected with ArT\'eMiS. The average projected separation between the LABOCA clumps is $1.0\pm0.2$~pc. The interclump dust emission between clumps~A and B is detected at both ArT\'eMiS wavelengths. That part of the filament exhibits three dust emission peaks in the ArT\'eMiS 450~$\mu$m image, and interestingly, those positions appear to be associated with \textit{Spitzer} 24~$\mu$m sources. The northernmost of these 24~$\mu$m sources could be identified with a young stellar object (YSO) candidate from the Robitaille et al. (2008) catalogue of intrinsically red mid-IR sources. The middle one of the 24~$\mu$m sources lies about $1\farcs5$ away from a target position called G11.38+0.81 P3 in Feng et al. (2000; their Fig.~1). That source is not detected as a dust emission peak in our ArT\'eMiS 350~$\mu$m or LABOCA 870~$\mu$m images. The southernmost of the 24~$\mu$m sources is also a prominent 70~$\mu$m source and hence can be considered a candidate YSO embedded in the filament. The average separation between the 450~$\mu$m maxima along the G11.36+0.80 filament, including those seen between clumps~A and B, is $0.55\pm0.14$~pc.

The submillimetre clumps~A and B in G11.36+0.80 are associated with weak 70~$\mu$m sources, but no counterparts were found from the \textit{Spitzer} 24~$\mu$m catalogue (\cite{gutermuth2015}). However, visual inspection of the \textit{Spitzer} 24~$\mu$m image suggests that some 24~$\mu$m emission is present in these clumps. Clumps~C and D are both associated with bright 24~$\mu$m sources and are hence likely to be in a more advanced stage of evolution than clumps~A and B (see Sect.~4.4. for further discussion). We note that the non-detection of a clump at 24~$\mu$m could be the result of a high column density of the source as well. However, the H$_2$ column densities towards clumps A--D are comparable with each other within the derived uncertainties (see column~(8) in Table~\ref{table:properties}).

The 24~$\mu$m image towards G11.36+0.80 (top middle panel in Fig.~\ref{figure:G1136images}) shows that there are dark absoprtion features to the south and south-east of clump~B. The features are partly dark even at 70~$\mu$m, and this region is coincident with a weak ($3\sigma$) protrusion-like feature in the LABOCA 870~$\mu$m image. These features could be subfilaments attached to G11.36+0.80. These types of subfilaments have been detected towards some other IRDCs as well (e.g. \cite{peretto2013}; \cite{miettinen2018}; \cite{leurini2019}), and they are denser structures than the striations seen to be connected to some lower-mass filaments (e.g. \cite{palmeirim2013}; \cite{bonne2020}; \cite{hacar2022} and \cite{pineda2022} for recent reviews). The subfilaments seen towards G11.36+0.80 could be signatures of the filament formation through gas flows or gas accretion from the surrounding medium. However, high-resolution interferometric molecular line observations would be needed to test these hypotheses and the physical connection of the features with the main filament in the first place. 

\begin{table}[H]
\renewcommand{\footnoterule}{}
\caption{Virial masses and virial parameters of the clumps in the target IRDCs.}
{\normalsize
\begin{minipage}{1\columnwidth}
\centering
\label{table:virial}
\begin{tabular}{c c c}
\hline\hline 
Source & $M_{\rm vir}$ & $\alpha_{\rm vir}$ \\
       & [M$_{\sun}$] &  \\ 
\hline 
\multicolumn{3}{c}{G1.75-0.08} \\
\hline
Clump A & $129\,337\pm13\,701$ & $23.51\pm4.71$ \\
Clump B & $36\,596\pm 2\,057$ & $9.71\pm1.98$ \\
\hline 
\multicolumn{3}{c}{G11.36+0.80} \\
\hline
Clump A & $226\pm6$ & $0.43\pm0.05$\\
Clump B & $151\pm2$ & $0.35\pm0.05$\\
Clump C & $280\pm24$ & $0.44\pm0.09$\\
Clump D & $101\pm3$ & $0.43\pm0.11$\\
\hline
\end{tabular} 
\end{minipage} }
\end{table}



\subsection{SED characteristics of clumps and filaments}

The physical properties of the clumps were derived using simple single- or two-temperature MBB models. The angular resolution of the \textit{Herschel}/PACS data at 70~$\mu$m and 160~$\mu$m, $5\farcs8\times 12\farcs1$ and $11\farcs4\times 13\farcs4$ (see Sect.~2.3), are comparable to those of the APEX bolometer data used in the SED fits, namely about $9\arcsec-19\farcs9$. Moreover, the PACS 160~$\mu$m and all the APEX bolometer flux densities were integrated within equal apertures, that is, within the region in which LABOCA emission was detected at $3\sigma$ significance. 

Figures~\ref{figure:G175_seds} and \ref{figure:G1136_seds} show that a single-temperature MBB \text{and a two-temperature MBB for 24~$\mu$m bright clumps} could be reasonably well fitted to the observed SEDs between 24~$\mu$m and 870~$\mu$m (the reduced $\chi^2$ values range from 0.90 to 3.55). Although we did not employ the \textit{Herschel}/SPIRE data in the fits, they are consistent with the SED fits for the clumps in G1.75-0.08 (especially the 250~$\mu$m flux density). The ArT\'eMiS 350~$\mu$m flux densities of the clumps in G1.75-0.08 (as integrated within the LABOCA $3\sigma$ contour) are $1.5\pm0.5$ (clump~A) and $1.2\pm0.4$ (clump~B) times the SPIRE 350~$\mu$m catalogue (\cite{molinari2016a}) flux densities, and hence agree fairly well within the uncertainties. For the G11.36+0.80 clumps, the SPIRE catalogue (\cite{molinari2016a}) flux densities show more scatter with respect to our SEDs, although in some cases, the agreement between the two is excellent (e.g. the 250~$\mu$m data point for clump~B (factor of 1.13 difference), and the SPIRE 350~$\mu$m data points for clumps~C and D (factors of 1.09 and 0.97 difference, respectively)). The ArT\'eMiS 350~$\mu$m flux densities of the clumps in G11.36+0.80 (within the LABOCA $3\sigma$ contour) are $0.7\pm0.2$ to $2.0\pm0.6$ times those of SPIRE at 350~$\mu$m. The best agreement between the two values is found for clump~C, where the ratio of the two flux density values is $1.0\pm0.3$.   

Figure~\ref{figure:G1136_seds} shows the SEDs of the G11.36+0.80 clumps~C and D would not be consistent with a single-temperature MBB model when the presence of 24~$\mu$m emission is taken into account. The 24~$\mu$m emission is arising from a warmer medium near the embedded YSO, while the longer-wavelength far-IR to submillimetre emission originates in a colder, dusty medium. Hence, a two-temperature MBB model was adopted for clumps~C and D to include the 24~$\mu$m flux density in the SED fit. This leads to a higher source luminosity than what would be derived from a single-$T$ model fit at $\lambda \geq 70$~$\mu$m as was done for clumps~A and B.

The SEDs of the target IRDC filaments in the wavelength range of 160--870~$\mu$m were also fitted with a single-temperature MBB model. In terms of the reduced $\chi^2$, the SED fit for G11.36+0.80 was a factor of 1.9 times better than for G1.75-0.08. For G1.75-0.08, the ArT\'eMiS flux densities lie above the model fit by factors of $1.9\pm0.6$ (350~$\mu$m) and $1.9\pm0.4$ (450~$\mu$m), while for G11.36+0.80 only the 450~$\mu$m data point lies above the MBB curve (by a factor of $1.6\pm0.3$). Within the flux density uncertainties, these differences are fairly small.

\subsection{Stability of the filaments and fragmentation into clumps}

\subsubsection{G1.75-0.08 -- a 70~$\mu$m dark binary clump system in the Galactic centre region}

The line mass (mass per unit length) of G1.75-0.08 is $M_{\rm line}=1\,011\pm146$~M$_{\sun}$~pc$^{-1}$. The dynamical state of the filament can be addressed by comparing its line mass with the virial or critical line mass (e.g. \cite{fiege2000}, Eq.~(12) therein). To calculate the latter quantity, we used the total (thermal+non-thermal) velocity dispersion, where the observed spectral line width (FWHM) was taken to be the FWHM of the HCN$(1-0)$ line detected towards clump~B in G1.75-0.08 from Miettinen (2014), that is, $13.50\pm0.38$~km~s$^{-1}$. We note that this is a very broad line width (and it is even broader (by a factor of 1.73) in clump~A; Sect.~3.5), which could be attributed to multiple factors. From an observational point of view, the hyperfine structure of HCN was not resolved in the MALT90 spectra, which can lead to an overestimation of the line width although it was derived through fitting the hyperfine structure of the transition (\cite{miettinen2014}). Moreover, the angular resolution of the Mopra telescope observations employed by Miettinen (2014) is $38\arcsec$ (HPBW), which corresponds to 1.5~pc at the cloud distance, and hence the beam might have captured emission from the turbulent outer parts of the cloud. For example, if G1.75-0.08 follows the line width–size relation in the central molecular zone, or CMZ, which for HCN is found to be $\sigma \propto R^{0.62}$ (\cite{shetty2012}; Table~2 therein), the aforementioned HCN$(1-0)$ line width would be expected to be only $\sim2.5$~km~s$^{-1}$ on the $\sim0.1$~pc scale, which is a typical inner width (FWHM) of filamentary molecular clouds (e.g. \cite{arzoumanian2019}; \cite{priestley2022}, and references therein). We note that Henshaw et al. (2016a) derived a velocity dispersion of 11~km~s$^{-1}$ ($\sim26$~km~s$^{-1}$ FWHM for a Gaussian profile) for another Galactic centre region IRDC, namely G0.253+0.016 or the Brick from spectral line observations with Mopra, which is comparable to the observational results for G1.75-0.08 with the same telescope (\cite{miettinen2014}). However, based on much higher angular resolution ($1\farcs7$) observations with the Atacama Large Millimetre/submillimetre Array (ALMA), Henshaw et al. (2019) derived an average velocity dispersion of 4.4~km~s$^{-1}$ in the Brick, which demonstrates that higher resolution spectral line observations towards G1.75-0.08 are also required. The effect of the angular resolution of the observations on the derived spectral line widths was also demonstrated by Hacar et al. (2018) in the case of the Orion integral filament (see e.g. Fig.~6 therein). From a physical point of view, there are several effects that can lead to line broadening. First, the HCN lines detected towards the G1.75-0.08 clumps were not optically thin, and hence the optical thickness effects might contribute to the broadening of the lines (e.g. \cite{hacar2016}). On the other hand, even the optically thin transitions detected towards the clumps had line widths that are comparable to those of HCN$(1-0)$ (\cite{miettinen2014}, Table~3 therein). Second, G1.75-0.08 might be associated with high-velocity gradients along the filament (e.g. \cite{federrath2016}; \cite{gong2018}) that would be blended in the MALT90 spectra and hence lead to overestimated line widths. Third, star formation driven outflows and shocks could lead to broad line profiles. Fourth, G1.75-0.08 is located close to the Galactic centre ($R_{\rm GC}\simeq270$~pc), where the interstellar medium is highly turbulent (e.g. \cite{salas2021} and references therein). At least part of this turbulent gas could contribute to our single-dish-observed spectral line widths. Obviously, further spectral line observations are needed to study the gas kinematics of G1.75-0.08 in more detail.

The gas kinetic temperature was assumed to be equal to the dust temperature ($15.0\pm0.4$~K) derived from the SED of the filament. For comparison, the isothermal sound speed at the cloud temperature is only $c_{\rm s}=0.2$~km~s$^{-1}$, and hence the filament appears to be dominated by supersonic non-thermal motions with $\sigma_{\rm NT}/c_{\rm s}\simeq 25$. The resulting virial line mass is 
$M_{\rm line}^{\rm vir}=15\,306\pm860$~M$_{\sun}$~pc$^{-1}$, and the corresponding $M_{\rm line}/M_{\rm line}^{\rm vir}$ ratio is about $0.07\pm0.01$. Hence, G1.75-0.08 as a whole appears to be gravitationally unbound, that is, subcritical. However, both $M_{\rm line}$ and $M_{\rm line}^{\rm vir}$ are particularly uncertain in the case of G1.75-0.08 owing to the uncertain dust opacity in the CMZ (e.g. owing to the change in metallicity) and the very large uncertainty in the gas velocity dispersion as discussed above. On the other hand, because G1.75-0.08 lies near the turbulent Galactic centre, its strongly subcritical (by a factor of $\sim14$) line mass could be the result of its environment.

Thermally supercritical filaments are generally found to be self-gravitating and in rough virial equilibrium (i.e. $M_{\rm line}\sim M_{\rm line}^{\rm vir}$ within a factor of $\sim2$; e.g. \cite{arzoumanian2013}; \cite{mattern2018b}). The derived $M_{\rm line}/M_{\rm line}^{\rm vir}$ ratio for G1.75-0.08 is very low compared to the ratio of the general population of Galactic filaments, but we emphasise again that our estimate of $M_{\rm line}^{\rm vir}$ is subject to significant uncertainties. Morever, at the small galactocentric distance of G1.75-0.08, the gas-to-dust mass ratio could be a factor of $\sim5$ lower than adopted in our study (\cite{giannetti2017}; Eq.~(2) therein), which would make the line mass of the filament lower by a similar factor (for a given dust opacity). On the other hand, a higher metallicity at small Galactocentric distances would make the mean molecular weight per free particle higher and hence the total velocity dispersion and virial line mass lower, which would increase the $M_{\rm line}/M_{\rm line}^{\rm vir}$ ratio. However, even if the metallicity were three times higher than assumed in the present work (\cite{giannetti2017}; Eq.~(3) therein), the mean molecular weight would increase by less than $\sim20\%$ (see \cite{kauffmann2008}, Appendix A.1 therein; for this estimate, the He mass fraction was assumed to increase with metallicity as $Y=0.24+2.4\times Z$ (e.g. \cite{mowlavi1998})). Hence, if the line mass of G1.75-0.08 is overestimated, which seems plausible owing to the Galactic gradient in the gas-to-dust ratio, the cloud could be even more subcritical and an even more severe outlier compared to the more general filamentary cloud population.

Another factor is that the line mass of G1.75-0.08 is not uniformly distributed, but the mass is concentrated at both ends of the filamentary structure. Hence, the system cannot be well approximated by idealised models of cylindrical cloud fragmentation. The theoretical clump separations calculated from multiple different cylindrical fragmentation models (\cite{larson1985}; \cite{nagasawa1987}; \cite{bastien1991}; \cite{jackson2010}; \cite{fischera2012}) were not found to agree well with the observed projected clump separation of 3.7~pc ($1.9\pm0.2$~pc if the 450~$\mu$m fragment in between clumps~A and B were counted as a third fragment). If the gas-to-dust mass ratio is overestimated, as discussed above, the cloud density needed in the fragmentation analyses would be different. Other caveats that complicate the comparison between the observed and theoretical clump separations is that the width of the G1.75-0.08 filament is fairly poorly defined as the system is composed of two large 
clumps that are connected by a narrower dust-emitting bridge, and the filament could be inclined from the line of sight.

A filament can also collapse along its long axis (longitudinal collapse) in such a way that the so-called gravitational focussing leads to the accumulation of gas at the ends of the filament, where gravitational collapse can then result in the formation of clumps (the so-called edge effect; e.g. \cite{burkert2004}; \cite{pon2012}; \cite{clarke2015}; \cite{heigl2021}; \cite{hoemann2022}). This configuration is qualitatively consistent with the configuration observed in G1.75-0.08. If the gravitational focussing scenario is true, the aforementioned possiblity of 
high-velocity gradients could well be present owing to the edge effect, where the filament self-gravity drives the piling up of gas at both ends of the filament. Owing to the cloud's location in the Galactic centre region, it might also be speculated that it was tidally stripped or disrupted, but the volume-averaged density of over $9\times10^3$~cm$^{-3}$ of the cloud makes this possibility questionable (\cite{kruijssen2014}). However, if the cloud mass is overestimated, then its true density would be lower and the cloud could be subject to the aforementioned disruption effects. The red-skewed profiles of HCN$(1-0)$ and HCO$^+(1-0)$ detected towards clump~A (\cite{miettinen2014}) are indeed an indication of outward motions. For comparison, orbital dynamics and shearing motions have been suggested to have influenced the structure and dynamics of the Brick (\cite{henshaw2019} and references therein), and these effects could also play a role in G1.75-0.08.

\subsubsection{Filamentary IRDC G11.36+0.80}

For G11.36+0.80, the line mass is $M_{\rm line}=376\pm59$~M$_{\sun}$~pc$^{-1}$. When we use the average FWHM of the detected N$_2$H$^+(1-0)$ lines (Table~\ref{table:line}, column~(3)), which is $1.39\pm0.13$~km~s$^{-1}$, the corresponding critical line mass becomes $M_{\rm line}^{\rm vir}=186\pm30$~M$_{\sun}$~pc$^{-1}$, which yields an $M_{\rm line}/M_{\rm line}^{\rm vir}$ ratio of $2.0\pm0.5$. Hence, G11.36+0.80 appears to be marginally supercritical. 

Miettinen (2012) concluded that G11.36+0.80 is fragmented into clumps as a result of the sausage instability. We can revisit this hypothesis using our improved estimates of the physical properties of the cloud. Our N$_2$H$^+(1-0)$ data suggest that the filament is moderately dominated by supersonic non-thermal motions ($\sigma_{\rm NT}/c_{\rm s}=2.6$ on average). First, we calculated the radial scale height of G11.36+0.80, which depends on the density at the centre of the filament (e.g. \cite{nagasawa1987}; Eq.~(2.3) therein). The average H$_2$ column density towards clumps~A--D is $\langle N({\rm H_2})\rangle=(2.0\pm0.3)\times10^{23}$~cm$^{-2}$ (Table~\ref{table:properties}, column~(8)). We can estimate the central H$_2$ number density by dividing the column density by the line-of-sight diameter of the filament, which we assumed to be equal to the width of the filament. This yields an average central H$_2$ number density of $(8.3\pm1.2)\times10^4$~cm$^{-3}$. The corresponding effective radial scale height of G11.36+0.80 is $H_{\rm eff}=0.04\pm0.01$~pc. This can be compared with the outer radius of the G11.36+0.80 filament as defined with respect to the LABOCA $3\sigma$ emission level, that is, 0.39~pc. Because the $R_{\rm fil}/H_{\rm eff}$ ratio is about $9.8\pm2.6$, G11.36+0.80 appears to be in a regime of $R \gg H$, where the fluid is compressional, and fragmentation owing to the sausage fluid instability is expected to produce fragments with a separation given by the wavelength of the fastest growing unstable mode, which is $22\times H_{\rm eff}$ (e.g. \cite{jackson2010}; \cite{anathpindika2021}). The theoretical clump separation, about $0.9\pm0.2$~pc, agrees very well with the observed LABOCA clump separation ($1.0\pm0.2$~pc on average). As seen in our 450~$\mu$m image, the filament region between clumps~A and B appears to be fragmented into substructures with an average separation of $0.19\pm0.08$~pc. This could be a manifestation of secular cloud fragmentation after the first density enhancements, the clumps seen in our LABOCA image, were formed via sausage instability (see Sect.~4.5).

A theoretical prediction for the maximum clump mass in G11.36+0.80 is about $22\times H_{\rm eff}\times M_{\rm line}$, which for the observed line mass is $331\pm98$~M$_{\sun}$. The most massive clump in G11.36+0.80 is derived to have a mass of $633\pm112$~M$_{\sun}$ (clump~C), which exceeds the aforementioned maximum mass by a factor of $1.9\pm0.7$. Hence, the two values are comparable within the uncertainties. We conclude that the mechanism by which G11.36+0.80 has fragmented into clumps can potentially be assigned to the sausage instability of the compressible fluid.

\subsection{Physical properties of the clumps and their potential for high-mass star formation}

The two clumps in G1.75-0.08 are found to be very massive ($\sim(3.8-5.5)\times10^3$~M$_{\sun}$), but their effective radii are also 
large ($\sim1.2-1.5$~pc within the LABOCA $3\sigma$ emission). As discussed in Sect.~4.1, neither of the clumps appears to have a 70~$\mu$m counterpart, and hence they are candidate high-mass starless clumps. However, because shock-tracer species such as SiO and HNCO have been detected in the G1.75-0.08 clumps (\cite{miettinen2014}), they might be associated with star formation activity, unless the shocks are caused by other processes as discussed above. Galactic clumps that appear dark at 70~$\mu$m are sometimes found to be associated with early stages of star formation (e.g. \cite{urquhart2022}). Because the aforementioned clumps do not appear to be gravitationally bound ($\alpha_{\rm vir}\gg2$; Table~\ref{table:virial}), they cannot be considered high-mass prestellar clumps even if they were starless. Moreover, as discussed in Sect.~4.3.1, the clump masses might be overestimated because the gas-to-dust ratio used in the calculation might have been too high, which would increase the virial parameter values. However, as also discussed in Sect.~4.3.1, the Mopra (MALT90) spectral line data used in the virial parameter calculation might overestimate the gas velocity dispersion compared to that within the clump interiors. For comparison, the MALT90 N$_2$H$^+(1-0)$ lines towards the LABOCA clumps in G11.36+0.80 are 1.25--1.43 (1.35 on average) times broader than the present N$_2$H$^+(1-0)$ lines observed with the IRAM 30-metre telescope. Nevertheless, the G1.75-0.08 clumps do appear to be dominated by the kinetic energy. That the clumps appear to be starless might be a result of turbulence, which is suggested to contribute to the suppression of star formation in the central part of the Galaxy (e.g. \cite{kruijssen2014}).

The clumps in G11.36+0.80 have masses in the range of $\sim 230-630$~M$_{\sun}$ and effective radii of $\sim0.4-0.6$~pc. All the clumps are found to be gravitationally bound ($\alpha_{\rm vir}<2$; Table~\ref{table:virial}), and the virial parameter is remarkably similar among the clumps ($\alpha_{\rm vir}\simeq0.4$). We note that Feng et al. (2020) derived SED-based dust temperatures for clumps~A and B (their sources P1 and P2) that agree well with our values (the ratio of the two is $0.85\pm0.01$ and $0.95\pm0.02$). Further literature comparison of the physical properties of the clumps is provided in Appendix~B.

The luminosity-to-mass ratios of clumps~A--D were derived to be about $0.4\pm0.1$, $0.4\pm0.1$, $0.5\pm0.1$, and $1.0\pm0.4$~L$_{\sun}$/M$_{\sun}$, respectively (the uncertainties were propagated from both quantities, and a mean value of the $\pm$ uncertainties in $L$ was used in the calculation). Because the $L/M$ ratio is expected to increase as the clump evolves, it can be used as an indicator of the clump evolutionary stage (e.g. \cite{ma2013}; \cite{molinari2016b}; \cite{elia2017}). We note that a similar evolutionary indicator (bolometric luminosity-to-envelope mass ratio, $L_{\rm bol}/M_{\rm env}$) is also used for low-mass -- class 0 and class I -- embedded protostars (e.g. \cite{andre2000} for a review). On the basis of this, clumps~A and B appear to be the least evolved sources in the filament. Clump~B in the centre of the filament is also associated with only a weak 70~$\mu$m source. Clumps~C and D appear to be somewhat more evolved sources owing to their slightly higher $L/M$ ratio (the differences in $L/M$ are not significant within the uncertainties). Clumps~C and D are both associated with 24~$\mu$m sources, which also suggests that the clumps are hosting more evolved embedded YSOs. 

\begin{figure}[!htb]
\centering
\resizebox{\hsize}{!}{\includegraphics{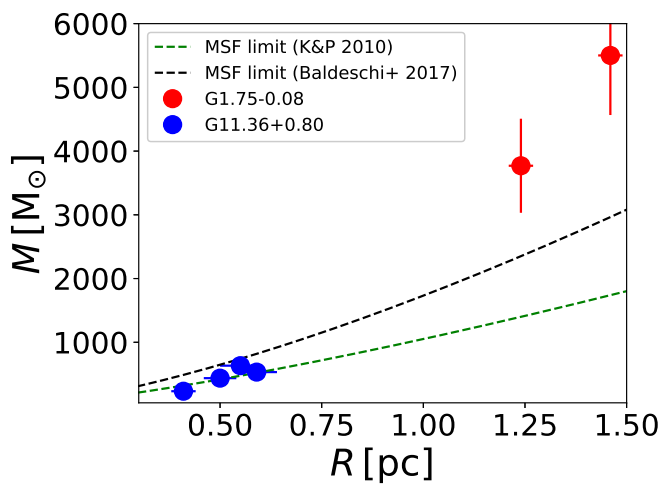}}
\caption{Clump mass plotted against the effective clump radius. The dashed green and black lines indicate the empirical mass-radius thresholds for high-mass star formation proposed by Kauffmann \& Pillai (2010) and Baldeschi et al. (2017), respectively.}
\label{figure:massvsradius}
\end{figure}

In Fig.~\ref{figure:massvsradius} we plot the masses of the studied clumps against their radius. For comparison, we also plot the mass-radius thresholds for high-mass star formation derived by Kauffmann \& Pillai (2010) and Baldeschi et al. (2017). The thresholds were scaled to the assumptions we adopted here (see Eqs.~(9) and (10) in Miettinen (2020)). The G1.75-0.08 clumps lie above the thresholds, for example by factors of $1.9\pm0.3$ (clump~A) and $1.6\pm0.3$ (clump~B) above the Baldeschi et al. (2017) limit. However, the clumps appear to be unbound by their self-gravity, and hence it is unclear whether they could collapse to form massive stars (see also Sect.~4.5). 

The G11.36+0.80 clumps lie very close to the Kauffmann \& Pillai (2010) threshold, that is, within factors of $0.7\pm0.2$ to $1.2\pm0.1$ ($1.0\pm0.1$ on average). Assuming a Kroupa (2001) stellar initial mass function and a star formation efficiency of 0.3, clumps~A--D could form a $14\pm2$~M$_{\sun}$, $12\pm1$~M$_{\sun}$, $15\pm1$~M$_{\sun}$, and a $7\pm0.4$~M$_{\sun}$ star, respectively (see \cite{svoboda2016}; \cite{sanhueza2017}). On the basis of this analysis, we cannot draw firm conclusions about the potential of the clumps in G11.36+0.80 to form massive stars, but clumps~A, B, and C could do so if the clump collapsed to form only one (massive) star or if the clump or cores within it accreted more gas from the parent filament.

\subsection{Evidence for hierarchical fragmentation}

\begin{figure*}[!htb]
\begin{center}
\includegraphics[width=0.315\textwidth]{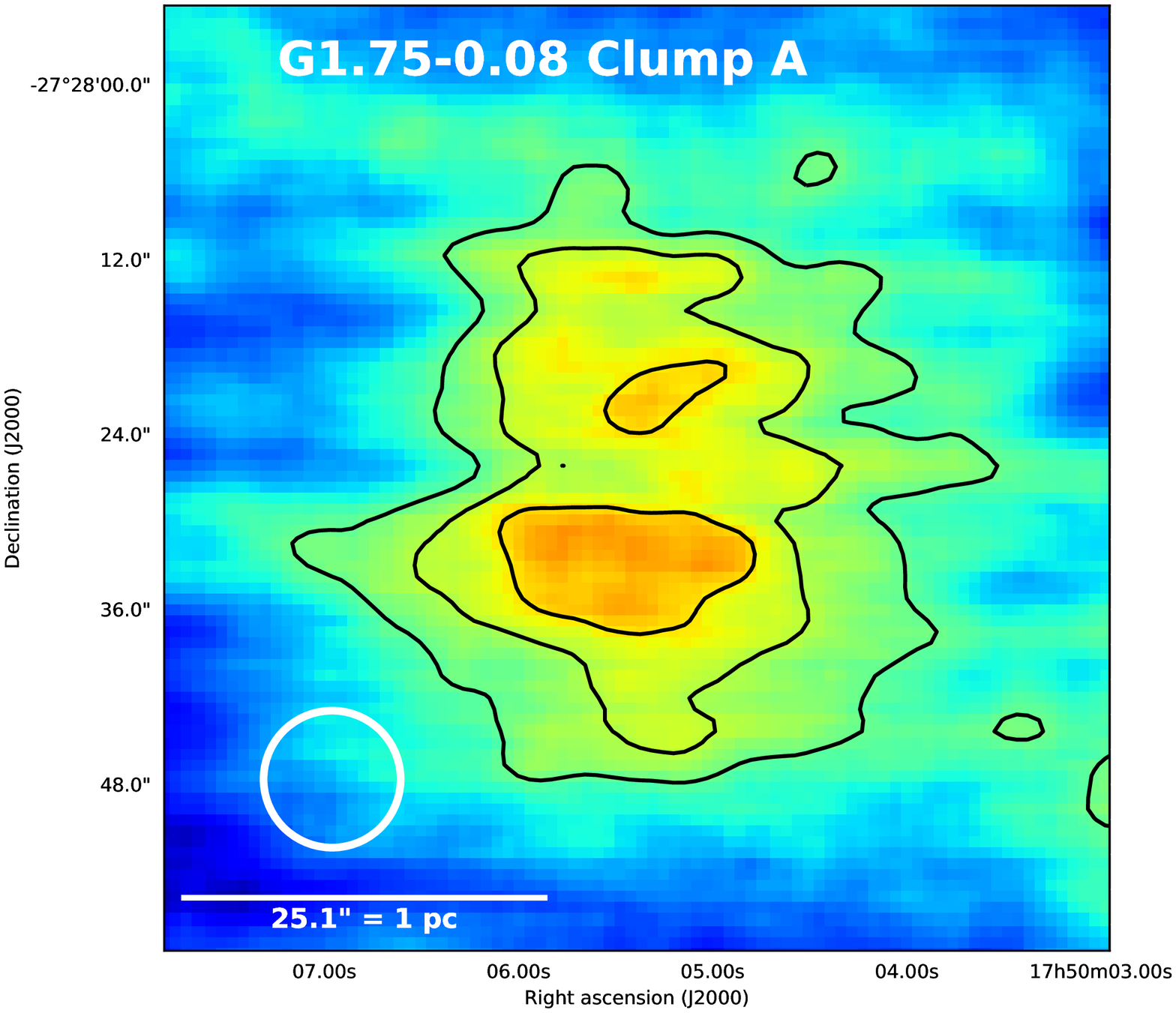}
\includegraphics[width=0.315\textwidth]{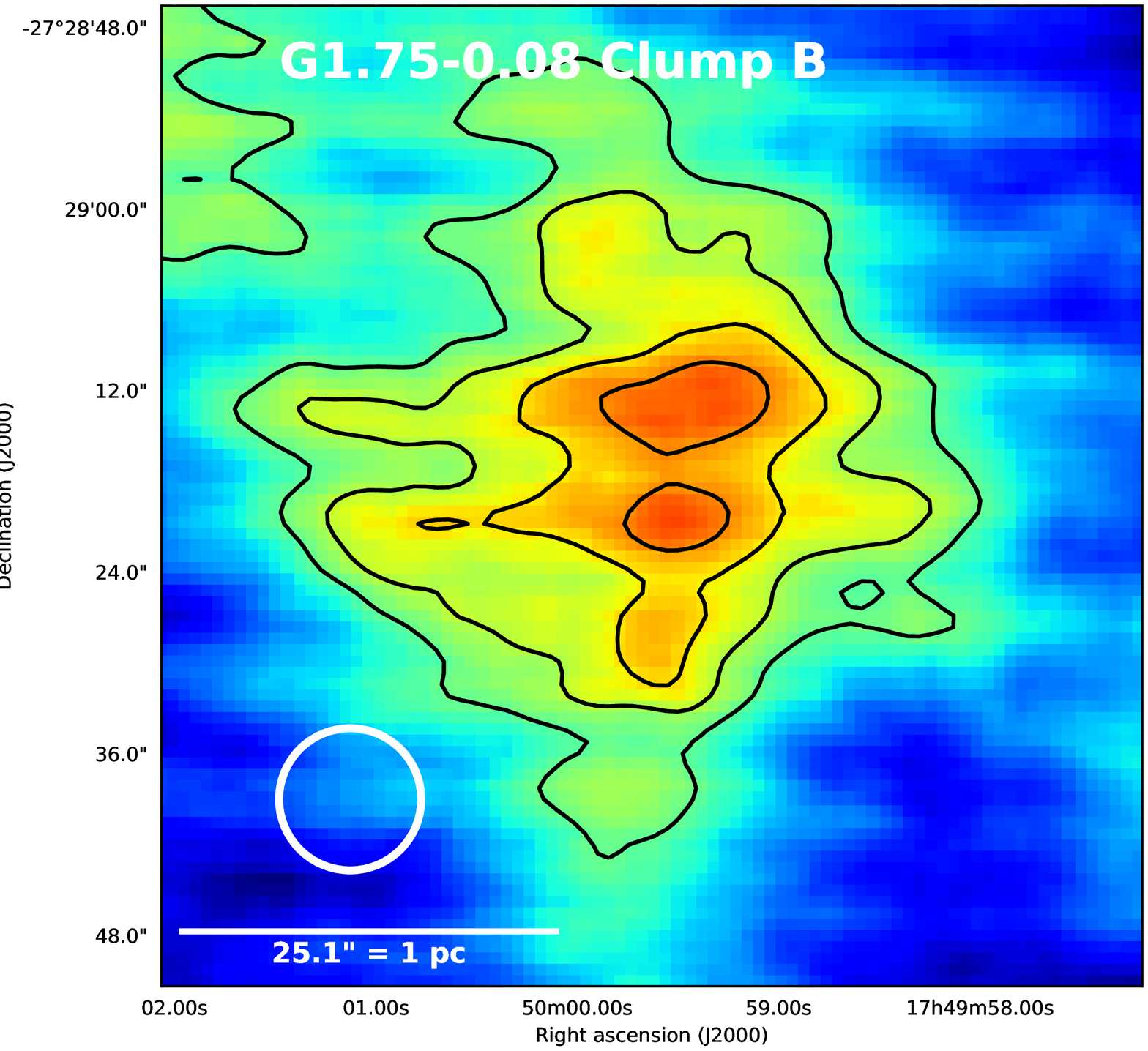}
\includegraphics[width=0.315\textwidth]{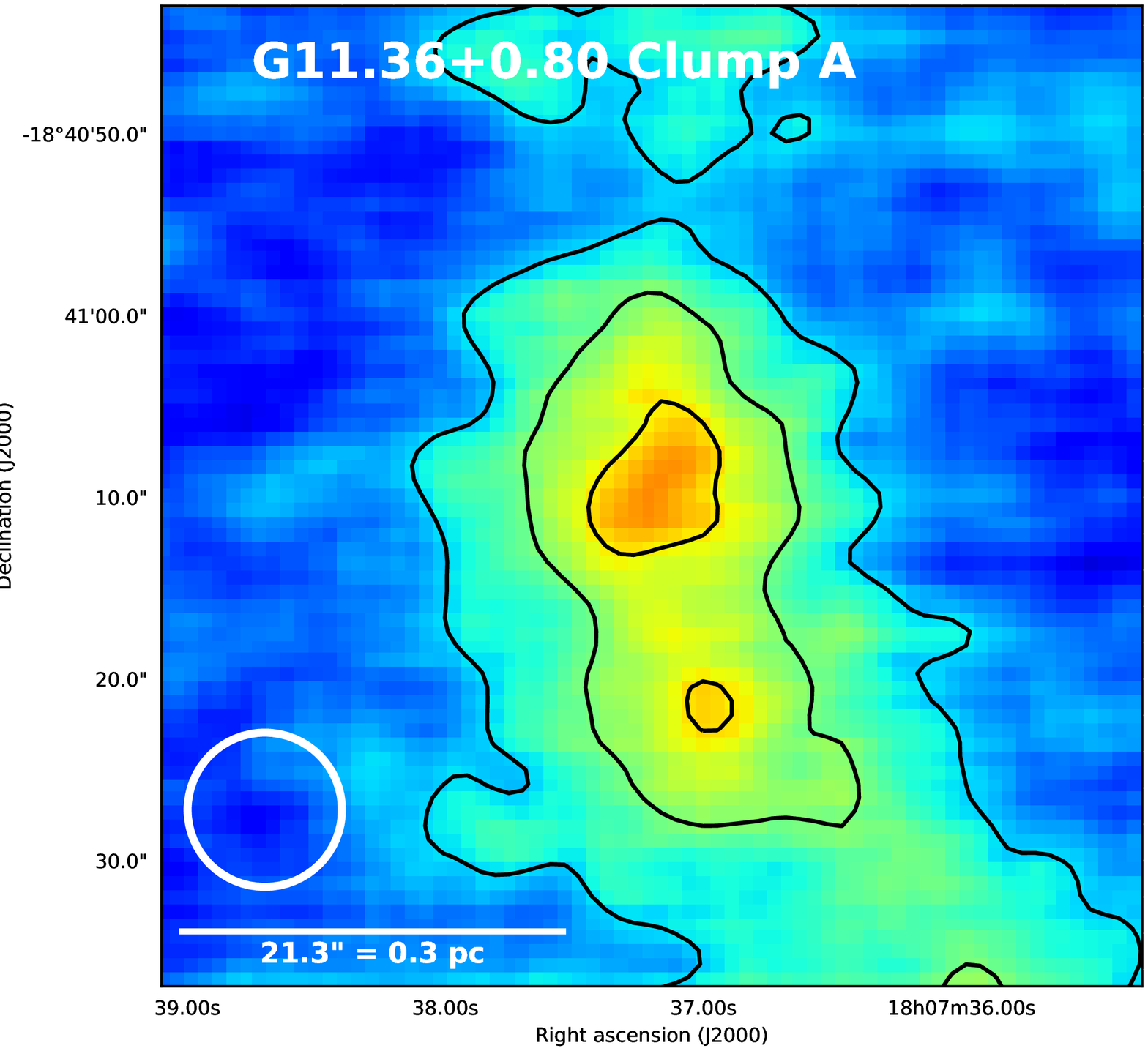}
\caption{Zoom-in images towards the clumps in our sample that show hints of substructures. ArT\'eMiS 450~$\mu$m images (colour scale and contours) are shown for the G1.75-0.08 clumps, while the ArT\'eMiS 350~$\mu$m image (colour scale and contours) is shown towards clump~A in G11.36+0.80. All the contours are as in Figs.~\ref{figure:G175images} and \ref{figure:G1136images}. The beam sizes (HPBW) are shown above the scale bars in each panel.}
\label{figure:zoomins}
\end{center}
\end{figure*}

In Fig.~\ref{figure:zoomins} we show the ArT\'eMiS images towards the clumps in our target IRDCs that show the clearest evidence of fragmention into smaller objects. In Table~\ref{table:fragmentation} we list the projected substructure separations in the clumps (column~(2)) and the thermal Jeans length of the clumps (column~(3); e.g. \cite{mckee2007} for a review, Eq.~(20) therein). The latter parameters agree within factors of 1.3--1.5 (column~(4) in Table~\ref{table:fragmentation}). The thermal Jeans masses of the clumps are listed in column~(5) in Table~\ref{table:fragmentation}. The masses of the substructures in the G1.75-0.08 clumps calculated by assuming that their flux density is equal to the ArT\'eMiS 450~$\mu$m peak intensity and that the dust temperature is equal to that of the parent clump (see e.g. Eq.~(2) in Miettinen et al. (2009)) are $\sim10^2$ times higher than the thermal Jeans mass. For clump~A in G11.36+0.80, the substructure masses are estimated to be $\sim10$ times higher than the thermal Jeans mass (column~(6) in Table~\ref{table:fragmentation}). However, the substructure mass estimates are subject to significant uncertainties owing to the uncertain dust properties (dust opacity and temperature and the dust-to-gas mass ratio). We note that the Jeans lengths that take the non-thermal motions into account would be clearly larger, by factors of $\sim4-38$ compared to the observed substructure separations. Although further observations are needed to study the clump fragmentation mechanisms in more detail, at least clump~A in G11.36+0.80 might have fragmented via thermal Jeans-like instability. 

Figure~\ref{figure:zoomins} shows that there are stripes in the ArT\'eMiS images along the right ascension axis that are artefacts owing to the scanning in only one direction. In principle, the detected substructure in the clumps could be the result of such artefacts, or at least the scanning direction could affect the observed morphology of the substructures (e.g. the southern substructure in clump~A of G1.75-0.08 is elongated in the right ascension direction). Moreover, the G1.75-0.08 clumps were found to be associated with strongly supersonic non-thermal motions ($\sigma_{\rm NT}/c_{\rm s}\simeq26-58$) and be gravitationally unbound, and hence the presence of substructure and Jeans fragmentation of the clumps might be questioned. However, Walker et al. (2021) found evidence that the fragments they detected in the Brick with ALMA at $0\farcs13$ resolution are the result of thermal Jeans fragmentation, although the parent cloud is highly turbulent. A similar conclusion was reached by Lu et al. (2020) on the basis of $\sim0\farcs2$ resolution ALMA observations of a sample of four molecular clouds in the Galactic centre region. 

Our findings for G11.36+0.80 are consistent with the paradigm where Galactic filaments fragment in a hierarchical, or at least in a two-level way, that is, the parent filament is fragmented into clump-size structures via cylindrical instability, and Jeans fragmentation takes place within the clumps at a higher density. As discussed in Sect.~1, this behaviour has been found in a wide variety of Galactic filaments, and the case of G11.36+0.80 with a filament mass of $\sim1.5\times10^3$~M$_{\sun}$ and high density of $\sim10^4$~cm$^{-3}$ adds another supportive data point to the growing body of evidence for a filamentary paradigm for star formation.

As discussed in Sect.~4.4, clump~A in G11.36+0.80 could in principle be able to form a high-mass star. However, as the clump appears to be fragmented into smaller substructures, it could represent the birth site of a stellar cluster of multiple low- to intermediate-mass stars. Massive star formation is still possible in this case via competitive accretion, where the gravitational potential well at the centre of the system allows for an enhanced gas accretion onto YSOs, and hence to reach high stellar masses (e.g. \cite{bonnell2006}). 

\begin{table*}
\caption{Clump substructure separations and the Jeans analysis parameters.}
\begin{minipage}{2\columnwidth}
\centering
\renewcommand{\footnoterule}{}
\label{table:fragmentation}
\begin{tabular}{c c c c c c c}
\hline\hline
Source & $d_{\rm sep}$ & $\lambda_{\rm J}$ & $d_{\rm sep}/\lambda_{\rm J}$ & $M_{\rm J}$ & $M_{\rm frag}/M_{\rm J}$\\
       & [pc] & [pc] &  & [M$_{\sun}$] &\\
\hline 
G1.75-0.08 clump A & $0.44\pm0.01$ & $0.29\pm0.02$ & $1.5\pm0.1$ & $5.6\pm1.7$ & $102\pm38$ and $61\pm28$\\
G1.75-0.08 clump B & $0.37\pm0.01$ & $0.28\pm0.03$ & $1.3\pm0.1$ & $5.4\pm1.9$ & $117\pm48$ and $115\pm50$\\
G11.36+0.80 clump A & $0.17\pm0.01$ & $0.24\pm0.01$ & $0.7\pm0.03$ & $4.3\pm0.8$ & $14\pm5$ and $7\pm3$\\
\hline
\end{tabular} 
\tablefoot{The listed parameters are the observed, projected separation between the substructures ($d_{\rm sep}$), thermal Jeans length ($\lambda_{\rm J}$), ratio between the latter two parameters ($d_{\rm sep}/\lambda_{\rm J}$), thermal Jeans mass ($M_{\rm J}$), and the ratio of the substructure mass to Jeans mass ($M_{\rm frag}/M_{\rm J}$; values are given for the two substructures in the clump).}
\end{minipage} 
\end{table*}

\section{Summary and conclusions}

We used ArT\'eMiS to image the IRDCs G1.75-0.08 and G11.36+0.80 at 350~$\mu$m and 450~$\mu$m. These new data were used in conjunction with our previous LABOCA 870~$\mu$m data on the IRDCs and archival \textit{Herschel} far-IR data. Spectral line observations of N$_2$H$^+(1-0)$ were also carried out towards the LABOCA 870~$\mu$m clumps in G11.36+0.80 with the IRAM 30~m telescope. Our main results are summarised as follows:

\begin{enumerate}
\item The IRDC G1.75-0.08 is composed of two cold ($T_{\rm dust}\sim14.5$~K), massive (several $10^3$~M$_{\sun}$) clumps that are projectively separated by $\sim3.7$~pc. Submillimetre dust emission is also detected in between the clumps, which in the LABOCA image appears to form a bridge-like structure connecting the clumps. Both the G1.75-0.08 clumps are 70~$\mu$m dark, and hence they could be massive starless clumps. Both clumps also fulfil the mass-radius threshold of high-mass star formation. However, they do not appear to be bounded by self-gravity and hence cannot be considered prestellar. Moreover, the ArT\'eMiS data suggest that the clumps have substructure. No firm conclusions about the cloud fragmentation mechanism can be drawn on the basis of the currently available data, but gravitational focussing could lead to a configuration in which a filament has clumps at both its ends. Owing to the cloud location near the Galactic centre ($\sim270$~pc), environmental conditions such as the elevated level of turbulence and tidal forces could affect the cloud dynamics. This hypothesis conforms to the finding that the cloud as whole appears strongly (by an order of magnitude) subcritical, which would make it an outlier compared to the general population of Galactic filaments. However, follow-up high-resolution spectral line observations are needed to confirm or disprove this.
\item The filamentary IRDC G11.36+0.80 contains four clear LABOCA 870~$\mu$m clumps that were all detected in the ArT\'eMiS images. The clumps have dust temperatures in the range $\sim13-15$~K and masses in the range $\sim 230-630$~M$_{\sun}$. The clumps are gravitationally bound, and they appear to be in somewhat different stages of temporal evolution on the basis of their luminosity-to-mass ratio. On the basis of the mass-radius comparison, the clumps do not appear to be  candidates for forming high-mass stars. Moreover, one of the G11.36+0.80 clumps (clump~A) shows signatures of further fragmentation into substructures in our 350~$\mu$m image. The G11.36+0.80 filament was found to be marginally supercritical (by a factor of two), and the projected separation of the LABOCA clumps is $\sim1$~pc on average. This periodic clump separation is consistent with a theoretical prediction of sausage instability in a compressible fluid. The observed substructure separation in clump~A is roughly consistent (within a factor of $\sim1.4$) with the thermal Jeans length of the clump, while the substructure masses exceed the corresponding thermal Jeans mass by about one order of magnitude. These findings suggest that the filament is fragmenting in a hierarchical way, but with different mechanisms on different spatial scales. 
\end{enumerate}

Our study is an example of a comparison of filamentary molecular clouds in two different Galactic environments.
Our results for G11.36+0.80 support the scenario in which filamentary IRDCs are fragmented via sausage instability, as has been proposed for some other IRDC filaments. A hierarchical or two-level fragmentation paradigm for Galactic filaments is also consistent with the detection of substructure in one of the G11.36+0.80 clumps. On the other hand, the physical characteristics and apparent lack of star formation activity in G1.75-0.08 might be the result of its location in the Galactic centre region, where dynamical effects such as tidal forces and shearing motions can affect the cloud evolution. Studies of larger samples of IRDCs at different galactocentric distances would be useful to quantify the role played by the cloud environment in its dynamics, fragmentation mechanism(s), and level of star formation, and to test the validity of the filament paradigm in the Galactic centre region. 

\begin{acknowledgements}

We thank the anonymous referee for providing extensive and insightful comments and suggestions that helped to improve the quality of this paper.
We are grateful to the staff at the APEX and IRAM 30~m telescopes for performing the service mode observations presented in this paper. M.~M. and Ph.~A. acknowledge support from “Ile de France” regional funding (DIM-ACAV+ Programme) and from the French national programmes of CNRS/INSU on 
stellar and ISM physics (PNPS and PCMI). This research has made use of NASA's Astrophysics Data System Bibliographic Services, the NASA/IPAC Infrared Science Archive, which is operated by the Jet Propulsion Laboratory, California Institute of Technology, under contract with the National Aeronautics and Space Administration, and {\tt Astropy}\footnote{\url{www.astropy.org}}, a community-developed core Python package for Astronomy (\cite{astropy2013}, \cite{astropy2018}). PACS has been developed by a consortium of institutes led by MPE (Germany) and including UVIE (Austria); KU Leuven, CSL, IMEC (Belgium); CEA, LAM (France); MPIA (Germany); INAF-IFSI/OAA/OAP/OAT, LENS, SISSA (Italy); IAC (Spain). This development has been supported by the funding agencies BMVIT (Austria), ESA-PRODEX (Belgium), CEA/CNES (France), DLR (Germany), ASI/INAF (Italy), and CICYT/MCYT (Spain). SPIRE has been developed by a consortium of institutes led by Cardiff University (UK) and including Univ. Lethbridge (Canada); NAOC (China); CEA, LAM (France); IFSI, Univ. Padua (Italy); IAC (Spain); Stockholm Observatory (Sweden); Imperial College London, RAL, UCL-MSSL, UKATC, Univ. Sussex (UK); and Caltech, JPL, NHSC, Univ. Colorado (USA). This development has been supported by national funding agencies: CSA (Canada); NAOC (China); CEA, CNES, CNRS (France); ASI (Italy); MCINN (Spain); SNSB (Sweden); STFC, UKSA (UK); and NASA (USA).

\end{acknowledgements}

\appendix

\section{\textit{Spitzer} and \textit{Herschel} counterparts of the detected sources}

To determine whether the submillimetre dust clumps in the target IRDCs are associated with YSOs, we cross-matched 
the ArT\'eMiS 350~$\mu$m peak positions of the clumps with the \textit{Spitzer}/MIPSGAL (\cite{carey2009}) catalogue (\cite{gutermuth2015}), which was accessed via the IRSA server. As a matching radius, we used a value of $10\arcsec$, which corresponds to about half the LABOCA beam size. Only two matches were found, namely for clumps C and D in G11.36+0.80 with angular separations between the 350~$\mu$m position and the MIPS 24~$\mu$m source of $8\farcs3$ and $5\farcs7$, respectively. 

The \textit{Herschel} source catalogues were also accessed via IRSA, and source counterparts could be found from the \textit{Herschel}/Hi-GAL (\cite{molinari2010}) catalogues (\cite{molinari2016a}). Again, we used the ArT\'eMiS 350~$\mu$m peak positions as the reference positions and used a search radius of $20\arcsec$, which is comparable to the LABOCA beam size. The 70~$\mu$m, 160~$\mu$m, 250~$\mu$m, 350~$\mu$m, and 500~$\mu$m counterparts were found within the projected distances of $4\farcs4-8\farcs7$, $0\farcs9-8\farcs2$, $1\farcs9-12\farcs7$, $1\farcs1-13\farcs2$, and $5\farcs0-16\farcs7$, respectively. Clump~D in G11.36+0.80 was an exception, where the SPIRE 500~$\mu$m counterpart position in the Hi-GAL catalogue is offset from the ArT\'eMiS 350~$\mu$m positions by $27\farcs1$, although the other \textit{Herschel} PACS and SPIRE positions lie only $4\farcs4-7\farcs0$ away from the ArT\'eMiS 350~$\mu$m emission maximum. Another special case is 
clump~B in G1.75-0.08, for which no PACS 160~$\mu$m counterpart could be found from the Hi-GAL catalogue, but visual inspection of the \textit{Herschel} 160~$\mu$m image (see Fig.~\ref{figure:G175images}, bottom left panel) clearly shows that the clump is seen in emission at this wavelength. Hence, we determined the corresponding flux density of the source from the 160~$\mu$m image using the nearby 160~$\mu$m counterpart of clump A as a benchmark to determine the same flux conversion factor (including the zero-level offset and background contribution) for clump B. We note that because clumps~A and B are projectively separated by only about 3.7~pc, the background (and foreground) contribution to their flux densities can be expected to be similar.

The photometric data of the \textit{Spitzer}/MIPS and \textit{Herschel}/PACS and SPIRE counterparts of the clumps in G1.75-0.08 and G11.36+0.80 are given in Table~\ref{table:photometry}.



\begin{table*}
\caption{Photometric data of the \textit{Spitzer}/MIPS 24~$\mu$m and \textit{Herschel}/PACS (70~$\mu$m and 160~$\mu$m) and SPIRE (250~$\mu$m, 350~$\mu$m, and 500~$\mu$m) counterparts of the clumps in G1.75-0.08 and G11.36+0.80.}
\begin{minipage}{2\columnwidth}
\centering
\renewcommand{\footnoterule}{}
\label{table:photometry}
\begin{tabular}{c c c c c c c}
\hline\hline
Source & $S_{24}$ & $S_{70}$ & $S_{160}$ & $S_{250}$ & $S_{350}$ & $S_{500}$\\
       & [mJy] & [Jy] & [Jy] & [Jy] & [Jy] & [Jy]\\
\hline 
\multicolumn{7}{c}{G1.75-0.08} \\
\hline
Clump A & \ldots & \ldots & $17.77\pm1.09$ & $80.71\pm2.16$ & $64.36\pm2.67$ & $38.96\pm4.40$ \\
Clump B & \ldots & \ldots & \ldots\tablefootmark{a} & $60.41\pm1.40$ & $51.52\pm2.15$ & $29.15\pm2.11$ \\
\hline 
\multicolumn{7}{c}{G11.36+0.80} \\
\hline
Clump A & \ldots & $0.38\pm0.02$ & $2.15\pm0.18$ & $14.04\pm0.43$ & $16.69\pm0.79$ & $25.80\pm1.38$ \\
Clump B & \ldots & $0.22\pm0.03$ & $1.26\pm0.19$ & $26.39\pm2.15$ & $16.01\pm0.57$ & $23.04\pm0.79$ \\
Clump C & $246.61\pm4.64$ & $6.44\pm0.09$ & $17.51\pm0.31$ & $28.29\pm0.64$ & $26.95\pm0.71$ & $28.81\pm1.56$\\
Clump D & $78.63\pm1.79$ & $4.51\pm0.08$ & $16.43\pm0.26$ & $13.77\pm0.39$ & $17.85\pm0.74$ & $10.13\pm1.74$\\
\hline
\end{tabular} 
\tablefoot{The flux densities were taken from the MIPSGAL and Hi-GAL catalogues unless otherwise stated.\tablefoottext{a}{No counterpart was found from the Hi-GAL catalogue. If the 160~$\mu$m flux density is determined from the PACS image within an aperture that corresponds to the $3\sigma$ level of the LABOCA emission, the value is $S_{160}=56.1\pm3.5$~Jy.}}
\end{minipage} 
\end{table*}

\section{Comparison with the $360\degr$ Hi-GAL catalogue of the physical properties of the clumps}

We cross-matched our clumps with the high-reliability version of the $360\degr$ Hi-GAL catalogue of the physical properties of the clumps (\cite{elia2021}). The cross-matching was performed with respect to the LABOCA peak positions of our clumps, and the search radius was set to $10\arcsec$, that is, about half the LABOCA beam size. All of our clumps were found to have a Hi-GAL counterpart, and they are listed in Table~\ref{table:elia} together with the physical properties as given in the Elia et al. (2021) catalogue.

Elia et al. (2021) derived the dust temperatures and source masses through fitting the source SEDs with a single-temperature MBB model at wavelengths $\lambda \geq 160$~$\mu$m. They assumed that the dust opacity is 10~cm$^2$~g$^{-1}$ at 300~$\mu$m, that the dust emissivity index is $\beta=2$, and that the gas-to-dust mass ratio is 100 (see \cite{elia2017}). Hence, the masses from Elia et al. (2021) should be multiplied by a factor of 1.216 to take the different assumptions into account when comparing with our results (and additionally to take the different source distances into account). The bolometric luminosities in the Elia et al. (2021) catalogue were calculated by integrating the SEDs from $\lambda=21$~$\mu$m to $\lambda=1.1$~mm, and the luminosities are proportional to the square of the distance, which also needs to be taken into account when their values are compared with our results. We provide below a source-by-source comparison of the physical properties we derived with those presented in the Hi-GAL catalogue of Elia et al. (2021).

The FWHM size we derived for clump~A in G1.75-0.08 from our ArT\'eMiS image at 350~$\mu$m ($44\farcs7$ when deconvolved from the beam size) 
is very similar (only 4\% difference) to that in the Hi-GAL catalogue (we note that 350~$\mu$m is the wavelength band closest to the 250~$\mu$m band adopted for the catalogue source sizes). No distance value for the clump was available in the catalogue, but if the 1~kpc based mass is scaled to 8.22~kpc and multiplied by the aforementioned correction factor, the value ($6\,006\pm1\,840$~M$_{\sun}$) becomes only $1.1\pm0.4$ times higher than our value. The dust temperature we derived for clump~A is about $1.1\pm0.1$ times the catalogue value and therefore agrees excellently. The bolometric luminosity we derived is 1.78 times higher than the scaled value from the catalogue, but we note that our luminosity value is associated with significant uncertainties (see Table~\ref{table:properties}, Col.~(7)).

For clump~B in G1.75-0.08, our deconvolved FWHM size agrees excellently with the catalogue value (the difference is only 1.7\%). A distance value was available in the Hi-GAL catalogue for this clump, and it is only 1.025 times lower than ours. The scaled value of the catalogue mass is $1.1\pm0.2$ times higher than our estimate, and the two values agree very well within the uncertainties. The dust temperatures also agree well within the uncertainties (our value is $1.2\pm0.1$ times higher). Our luminosity is 2.2 times higher, but again, we note that our value is associated with large uncertainties. We note that the source distances in the Elia et al. (2021) catalogue were taken from M{\`e}ge et al. (2021), and although the reported distance is comparable to our value, it was derived from an LSR velocity of 117.2~km~s$^{-1}$, which is completely different from the LSR velocities of the MALT90 spectral line detections (around 50~km~s$^{-1}$; \cite{miettinen2014}). We also note that the M{\`e}ge et al. (2021) Hi-GAL catalogue does not contain data on the velocity dispersion that we could compare with our line width values for the G1.75-0.08 filament.

Only clumps~B and D in G11.36+0.80 had distance values in the Hi-GAL catalogue, and only for clump~D this value is similar to ours (the difference is only a factor of 1.04). The catalogue distance of clump~B is 5.5~kpc, but our N$_2$H$^+(1-0)$ observations show that the clump is physically associated with the filament at 3.23~kpc. Our deconvolved FWHM sizes at 350~$\mu$m for the clumps in G11.36+0.80 are 1.17 to 3.51 times larger (2.17 on average) than the catalogue values. Especially the reported size of clump~D in the catalogue is very compact, only $3\farcs42$, which is about five times smaller than the SPIRE beam at 250~$\mu$m. Our nominal clump masses are about 0.4 to 5.1 times the catalogue values (2.4 times higher on average). The best agreement is found for clump~B, where our value is $1.8\pm0.3$ times the catalogue value. The poorest agreement is with clump~D, for which our mass estimate is $5.1\pm1.2$ times higher. However, for clump~D, we used a two-temperature MBB SED fitting approach, and hence a direct comparison with the Hi-GAL catalogue value is not straightforward and our higher mass estimate is not surprising. The dust temperatures we derived are $0.7-1.5$ (1.04 on average) times the catalogue values. The agreement is excellent especially for clumps~B and C,  although for clump~C, we used a two-temperature MBB fit. The luminosities we derived are $1.1-3.5$ times higher (1.9 on average) than those in the catalogue, but owing to the large uncertainties in our derived luminosities, these differences are not significant. Overall, the physical properties we derived for our clumps agree well with those reported in the $360\degr$ Hi-GAL catalogue of Elia et al. (2021).

\begin{table*}
\caption{Source counterparts in the $360\degr$ Hi-GAL catalogue of clump physical properties (\cite{elia2021}).}
\begin{minipage}{2\columnwidth}
\centering
\renewcommand{\footnoterule}{}
\label{table:elia}
\begin{tabular}{c c c c c c c c}
\hline\hline
Source & Hi-GAL designation & Sep. & $d$ & FWHM & $M$ & $T_{\rm dust}$ & $L_{\rm bol}$\\
       &  & [$\arcsec$] & [kpc] & [$\arcsec$] & [M$_{\sun}$] & [K] & [L$_{\sun}$]\\
\hline 
\multicolumn{8}{c}{G1.75-0.08} \\
\hline
Clump A & HIGALBM1.7613-0.0857 & 9.94 & \ldots & 42.81 & $73.1\pm22.4$ & $12.6\pm1.3$ & 26.1\\
Clump B & HIGALBM1.7361-0.0762 & 2.87 & 8.02 & 37.67 & $3\,202.2\pm9.2$ & $12.3\pm0.5$ & 1\,028.5\\
\hline 
\multicolumn{8}{c}{G11.36+0.80} \\
\hline
Clump A & HIGALBM11.3848+0.8110 & 5.27 & \ldots & 18.79 & $96.5\pm1.1$ & $9.0\pm0.0$ & 6.1\\
Clump B & HIGALBM11.3618+0.8016 & 4.93 & 5.5 & 9.26 & $572.4\pm0.5$ & $13.1\pm0.2$ & 295.8\\
Clump C & HIGALBM11.3449+0.7973 & 8.47 & \ldots & 13.62 & $22.0\pm0.5$ & $13.4\pm0.1$ & 25.6\\
Clump D & HIGALBM11.3360+0.7853 & 3.66 & 3.37 & 3.42 & $40.7\pm0.3$ & $21.4\pm0.4$ & 257.2\\
\hline
\end{tabular} 
\tablefoot{The columns are as follows: (1) source name; (2) source designation in the Elia et al. (2021) Hi-GAL catalogue; (3) projected angular separation between the LABOCA 870~$\mu$m peak position of our source and the Hi-GAL catalogue position; (4) source distance in the Hi-GAL catalogue (not available for clump~A in G1.75-0.08 and clumps~A and C in G11.36+0.80); (5) circularised and beam-deconvolved FWHM size measured from the \textit{Herschel}/SPIRE 250~$\mu$m image; (6) total clump mass derived from fitting the source SED, and where a hypothetical distance of 1~kpc was used for sources with unavailable distance; (7) dust temperature derived from the SED fit; and (8) bolometric luminosity derived by integrating the SED from $\lambda=21$~$\mu$m to $\lambda=1.1$~mm, where again a distance of 1~kpc was assumed if it was not available.}
\end{minipage} 
\end{table*}


\begin{thebibliography}{}

\bibitem[Anathpindika \& Francesco 2021]{anathpindika2021} Anathpindika, S.~V. \& Francesco, J.~D.\ 2021, \mnras, 502, 564

\bibitem[Andr{\'e} et al. 2000]{andre2000} Andr{\'e}, P., Ward-Thompson, D., \& Barsony, M.\ 2000, in \textit{Protostars and Planets IV}, V. Mannings, A. P. Boss, and S. S. Russell (eds.), University of Arizona Press, Tucson, p.~59

\bibitem[Andr{\'e} et al. 2014]{andre2014} Andr{\'e}, P., Di Francesco, J., Ward-Thompson, D., et al.\ 2014, in \textit{Protostars and Planets VI}, H. Beuther, R. S. Klessen, C. P. Dullemond, and Th. Henning (eds.), University of Arizona Press, Tucson, 914 pp., p.~27--51

\bibitem[Andr{\'e} et al. 2016]{andre2016} Andr{\'e}, P., Rev{\'e}ret, V., K{\"o}nyves, V., et al.\ 2016, \aap, 592, A54

\bibitem[Andr{\'e} et al. 2019]{andre2019} Andr{\'e}, P., Arzoumanian, D., K{\"o}nyves, V., et al.\ 2019, \aap, 629, L4

\bibitem[Arzoumanian et al. 2013]{arzoumanian2013} Arzoumanian, D., Andr{\'e}, P., Peretto, N., et al.\ 2013, \aap, 553, A119

\bibitem[Arzoumanian et al. 2019]{arzoumanian2019} Arzoumanian, D., Andr{\'e}, P., K{\"o}nyves, V., et al.\ 2019, \aap, 621, A42

\bibitem[Astropy Collaboration et al. 2013]{astropy2013} Astropy Collaboration, Robitaille, T.~P., Tollerud, E.~J., et al.\ 2013, \aap, 558, A33

\bibitem[2018]{astropy2018} Astropy Collaboration, Price-Whelan, A.~M., Sip{\H{o}}cz, B.~M., et al.\ 2018, \aj, 156, 123

\bibitem[Baldeschi et al. 2017]{baldeschi2017} Baldeschi, A., Elia, D., Molinari, S., et al.\ 2017, \mnras, 466, 3682

\bibitem[Bastien et al. 1991]{bastien1991} Bastien, P., Arcoragi, J.-P., Benz, W., et al.\ 1991, \apj, 378, 255

\bibitem[Battersby et al. 2010]{battersby2010} Battersby, C., Bally, J., Jackson, J.~M., et al.\ 2010, \apj, 721, 222

\bibitem[Beuther \& Steinacker 2007]{beuther2007} Beuther, H., \& Steinacker, J.\ 2007, \apjl, 656, L85

\bibitem[Bonne et al. 2020]{bonne2020} Bonne, L., Bontemps, S., Schneider, N., et al.\ 2020, \aap, 644, A27

\bibitem[Bonnell \& Bate 2006]{bonnell2006} Bonnell, I.~A. \& Bate, M.~R.\ 2006, \mnras, 370, 488

\bibitem[Burkert \& Hartmann 2004]{burkert2004} Burkert, A. \& Hartmann, L.\ 2004, \apj, 616, 288

\bibitem[Busquet et al. 2013]{busquet2013} Busquet, G., Zhang, Q., Palau, A., et al.\ 2013, \apjl, 764, L26

\bibitem[Carey et al. 2009]{carey2009} Carey, S.~J., Noriega-Crespo, A., Mizuno, D.~R., et al.\ 2009, \pasp, 121, 76

\bibitem[Carter et al. 2012]{carter2012} Carter, M., Lazareff, B., Maier, D., et al.\ 2012, \aap, 538, A89

\bibitem[Chambers et al. 2009]{chambers2009} Chambers, E.~T., Jackson, J.~M., Rathborne, J.~M., et al.\ 2009, \apjs, 181, 360

\bibitem[Churchwell et al. 2009]{churchwell2009} Churchwell, E., Babler, B.~L., Meade, M.~R., et al.\ 2009, \pasp, 121, 213


\bibitem[Clarke \& Whitworth 2015]{clarke2015} Clarke, S.~D. \& Whitworth, A.~P.\ 2015, \mnras, 449, 1819

\bibitem[Egan et al. 1998]{egan1998} Egan, M.~P., Shipman, R.~F., Price, S.~D., et al.\ 1998, \apjl, 494, L199

\bibitem[Elia et al. 2017]{elia2017} Elia, D., Molinari, S., Schisano, E., et al.\ 2017, \mnras, 471, 100

\bibitem[Elia et al. 2021]{elia2021} Elia, D., Merello, M., Molinari, S., et al.\ 2021, \mnras, 504, 2742

\bibitem[Fazio et al. 2004]{fazio2004} Fazio, G.~G., Hora, J.~L., Allen, L.~E., et al.\ 2004, \apjs, 154, 10.

\bibitem[Federrath et al. 2016]{federrath2016} Federrath, C., Rathborne, J.~M., Longmore, S.~N., et al.\ 2016, \apj, 832, 143

\bibitem[Feng et al. 2020]{feng2020} Feng, S., Li, D., Caselli, P., et al.\ 2020, \apj, 901, 145

\bibitem[Fiege \& Pudritz 2000]{fiege2000} Fiege, J.~D. \& Pudritz, R.~E.\ 2000, \mnras, 311, 85

\bibitem[Fischera \& Martin 2012]{fischera2012} Fischera, J. \& Martin, P.~G.\ 2012, \aap, 542, A77

\bibitem[Foster et al. 2011]{foster2011} Foster, J.~B., Jackson, J.~M., Barnes, P.~J., et al.\ 2011, \apjs, 197, 25

\bibitem[Foster et al. 2013]{foster2013} Foster, J.~B., Rathborne, J.~M., Sanhueza, P., et al.\ 2013, \pasa, 30, e038

\bibitem[Foster et al. 2014]{foster2014} Foster, J.~B., Arce, H.~G., Kassis, M., et al.\ 2014, \apj, 791, 108

\bibitem[Giannetti et al. 2017]{giannetti2017} Giannetti, A., Leurini, S., K{\"o}nig, C., et al.\ 2017, \aap, 606, L12

\bibitem[Gong et al. 2018]{gong2018} Gong, Y., Li, G.~X., Mao, R.~Q., et al.\ 2018, \aap, 620, A62

\bibitem[Griffin et al. 2010]{griffin2010} Griffin, M.~J., Abergel, A., Abreu, A., et al.\ 2010, \aap, 518, L3

\bibitem[Gutermuth \& Heyer 2015]{gutermuth2015} Gutermuth, R.~A. \& Heyer, M.\ 2015, \aj, 149, 64

\bibitem[G{\"u}sten et al. 2006]{gusten2006} G{\"u}sten, R., Nyman, L. {\r{A}}., Schilke, P., et al.\ 2006, \aap, 454, L13

\bibitem[Hacar et al. 2016]{hacar2016} Hacar, A., Alves, J., Burkert, A., et al.\ 2016, \aap, 591, A104

\bibitem[Hacar et al. 2018]{hacar2018} Hacar, A., Tafalla, M., Forbrich, J., et al.\ 2018, \aap, 610, A77

\bibitem[Hacar et al. 2022]{hacar2022} Hacar, A., Clark, S., Heitsch, F., et al.\ 2022, in \textit{Protostars and Planets VII}, S. Inutsuka, Y. Aikawa, T. Muto, K. Tomida, and M. Tamura (eds.), University of Arizona Press, Tucson, \textit{in press}, {\tt arXiv:2203.09562}

\bibitem[Heigl et al. 2021]{heigl2021} Heigl, S., Hoemann, E., \& Burkert, A.\ 2021, {\tt arXiv:2112.12640}

\bibitem[Henshaw et al. 2016a]{henshaw2016a} Henshaw, J.~D., Longmore, S.~N., Kruijssen, J.~M.~D., et al.\ 2016a, \mnras, 457, 2675

\bibitem[Henshaw et al. 2016b]{henshaw2016b} Henshaw, J.~D., Caselli, P., Fontani, F., et al.\ 2016b, \mnras, 463, 146

\bibitem[Henshaw et al. 2019]{henshaw2019} Henshaw, J.~D., Ginsburg, A., Haworth, T.~J., et al.\ 2019, \mnras, 485, 2457

\bibitem[Hoemann et al. 2022]{hoemann2022} Hoemann, E., Heigl, S., \& Burkert, A.\ 2022, \mnras, \textit{submitted}, {\tt arXiv:2203.07002}

\bibitem[Jackson et al. 2010]{jackson2010} Jackson, J.~M., Finn, S.~C., Chambers, E.~T., et al.\ 2010, \apjl, 719, L185

\bibitem[Jackson et al. 2013]{jackson2013} Jackson, J.~M., Rathborne, J.~M., Foster, J.~B., et al.\ 2013, \pasa, 30, e057

\bibitem[Kainulainen et al. 2013]{kainulainen2013} Kainulainen, J., Ragan, S.~E., Henning, T., et al.\ 2013, \aap, 557, A120

\bibitem[Kauffmann \& Pillai 2010]{kauffmann2010} Kauffmann, J. \& Pillai, T.\ 2010, \apjl, 723, L7

\bibitem[Kauffmann et al. 2008]{kauffmann2008} Kauffmann, J., Bertoldi, F., Bourke, T.~L., et al.\ 2008, \aap, 487, 993

\bibitem[Kroupa 2001]{kroupa2001} Kroupa, P.\ 2001, \mnras, 322, 231

\bibitem[Kruijssen et al. 2014]{kruijssen2014} Kruijssen, J.~M.~D., Longmore, S.~N., Elmegreen, B.~G., et al.\ 2014, \mnras, 440, 3370

\bibitem[Larson 1985]{larson1985} Larson, R.~B.\ 1985, \mnras, 214, 379

\bibitem[Leurini et al. 2019]{leurini2019} Leurini, S., Schisano, E., Pillai, T., et al.\ 2019, \aap, 621, A130

\bibitem[Liu et al. 2021]{liu2021} Liu, H.-L., Tej, A., Liu, T., et al.\ 2021, \mnras, \textit{in press}

\bibitem[Lu et al. 2020]{lu2020} Lu, X., Cheng, Y., Ginsburg, A., et al.\ 2020, \apjl, 894, L14

\bibitem[Ma et al. 2013]{ma2013} Ma, B., Tan, J.~C., \& Barnes, P.~J.\ 2013, \apj, 779, 79

\bibitem[Mangum \& Shirley 2015]{mangum2015} Mangum, J.~G. \& Shirley, Y.~L.\ 2015, \pasp, 127, 266


\bibitem[Mattern et al. 2018a]{mattern2018a} Mattern, M., Kainulainen, J., Zhang, M., et al.\ 2018a, \aap, 616, A78

\bibitem[Mattern et al. 2018b]{mattern2018b} Mattern, M., Kauffmann, J., Csengeri, T., et al.\ 2018b, \aap, 619, A166

\bibitem[McKee \& Ostriker 2007]{mckee2007} McKee, C.~F. \& Ostriker, E.~C.\ 2007, \araa, 45, 565

\bibitem[M{\`e}ge et al. 2021]{mege2021} M{\`e}ge, P., Russeil, D., Zavagno, A., et al.\ 2021, \aap, 646, A74

\bibitem[Miettinen 2012]{miettinen2012} Miettinen, O.\ 2012, \aap, 540, A104

\bibitem[Miettinen 2014]{miettinen2014} Miettinen, O.\ 2014, \aap, 562, A3

\bibitem[Miettinen 2018]{miettinen2018} Miettinen, O.\ 2018, \aap, 609, A123

\bibitem[Miettinen 2020]{miettinen2020} Miettinen, O.\ 2020, \aap, 644, A82

\bibitem[Miettinen et al. 2009]{miettinen2009} Miettinen, O., Harju, J., Haikala, L.~K., et al.\ 2009, \aap, 500, 845

\bibitem[Molinari et al. 2010]{molinari2010} Molinari, S., Swinyard, B., Bally, J., et al.\ 2010, \pasp, 122, 314

\bibitem[Molinari et al. 2016a]{molinari2016a} Molinari, S., Schisano, E., Elia, D., et al.\ 2016a, \aap, 591, A149

\bibitem[Molinari et al. 2016b]{molinari2016b} Molinari, S., Merello, M., Elia, D., et al.\ 2016b, \apjl, 826, L8

\bibitem[Motte et al. 2018]{motte2018} Motte, F., Bontemps, S., \& Louvet, F.\ 2018, \araa, 56, 41

\bibitem[Mowlavi et al. 1998]{mowlavi1998} Mowlavi, N., Meynet, G., Maeder, A., et al.\ 1998, \aap, 335, 573

\bibitem[Nagasawa 1987]{nagasawa1987} Nagasawa, M.\ 1987, Progress of Theoretical Physics, 77, 635

\bibitem[Ossenkopf \& Henning 1994]{ossenkopf1994} Ossenkopf, V. \& Henning, T.\ 1994, \aap, 291, 943

\bibitem[Pagani et al. 2009]{pagani2009} Pagani, L., Daniel, F., \& Dubernet, M.-L.\ 2009, \aap, 494, 719

\bibitem[Palmeirim et al. 2013]{palmeirim2013} Palmeirim, P., Andr{\'e}, P., Kirk, J., et al.\ 2013, \aap, 550, A38

\bibitem[P\'erault et al. 1996]{perault1996} P\'erault, M., Omont, A., Simon, G., et al.\ 1996, \aap, 315, L165

\bibitem[Peretto \& Fuller 2009]{peretto2009} Peretto, N., \& Fuller, G.~A.\ 2009, \aap, 505, 405

\bibitem[Peretto et al. 2013]{peretto2013} Peretto, N., Fuller, G.~A., Duarte-Cabral, A., et al.\ 2013, \aap, 555, A112

\bibitem[Pilbratt et al. 2010]{pilbratt2010} Pilbratt, G.~L., Riedinger, J.~R., Passvogel, T., et al.\ 2010, \aap, 518, L1

\bibitem[Pineda et al. 2022]{pineda2022} Pineda, J.~E., Arzoumanian, D., Andr{\'e}, P., et al.\ 2022, in \textit{Protostars and Planets VII}, S. Inutsuka, Y. Aikawa, T. Muto, K. Tomida, and M. Tamura (eds.), University of Arizona Press, Tucson, \textit{in press}, {\tt arXiv:2205.03935}

\bibitem[Poglitsch et al. 2010]{poglitsch2010} Poglitsch, A., Waelkens, C., Geis, N., et al.\ 2010, \aap, 518, L2

\bibitem[Pon et al. 2012]{pon2012} Pon, A., Toal{\'a}, J.~A., Johnstone, D., et al.\ 2012, \apj, 756, 145

\bibitem[Priestley \& Whitworth 2022]{priestley2022} Priestley, F.~D. \& Whitworth, A.~P.\ 2022, \mnras, 509, 1494

\bibitem[Rathborne et al. 2006]{rathborne2006} Rathborne, J.~M., Jackson, J.~M., \& Simon, R.\ 2006, \apj, 641, 389

\bibitem[Reid et al. 2014]{reid2014} Reid, M.~J., Menten, K.~M., Brunthaler, A., et al.\ 2014, \apj, 783, 130

\bibitem[Rev{\'e}ret et al. 2014]{reveret2014} Rev{\'e}ret, V., Andr{\'e}, P., Le Pennec, J., et al.\ 2014, \procspie, 9153, 915305

\bibitem[Rieke et al. 2004]{rieke2004} Rieke, G.~H., Young, E.~T., Engelbracht, C.~W., et al.\ 2004, \apjs, 154, 25

\bibitem[Robitaille et al. 2008]{robitaille2008} Robitaille, T.~P., Meade, M.~R., Babler, B.~L., et al.\ 2008, \aj, 136, 2413

\bibitem[Rosolowsky et al. 2008]{rosolowsky2008} Rosolowsky, E.~W., Pineda, J.~E., Kauffmann, J., et al.\ 2008, \apj, 679, 1338

\bibitem[Salas et al. 2021]{salas2021} Salas, J.~M., Morris, M.~R., \& Naoz, S.\ 2021, \aj, 161, 243.

\bibitem[Sanhueza et al. 2017]{sanhueza2017} Sanhueza, P., Jackson, J.~M., Zhang, Q., et al.\ 2017, \apj, 841, 97

\bibitem[Shetty et al. 2012]{shetty2012} Shetty, R., Beaumont, C.~N., Burton, M.~G., et al.\ 2012, \mnras, 425, 720

\bibitem[Shimajiri et al. 2019]{shimajiri2019} Shimajiri, Y., Andr{\'e}, P., Ntormousi, E., et al.\ 2019, \aap, 632, A83

\bibitem[Simon et al. 2006]{simon2006} Simon, R., Jackson, J.~M., Rathborne, J.~M., et al.\ 2006, \apj, 639, 227

\bibitem[Siringo et al. 2009]{siringo2009} Siringo, G., Kreysa, E., Kov{\'a}cs, A., et al.\ 2009, \aap, 497, 945

\bibitem[Svoboda et al. 2016]{svoboda2016} Svoboda, B.~E., Shirley, Y.~L., Battersby, C., et al.\ 2016, \apj, 822, 59

\bibitem[Takahashi et al. 2013]{takahashi2013} Takahashi, S., Ho, P.~T.~P., Teixeira, P.~S., et al.\ 2013, \apj, 763, 57

\bibitem[Talvard et al. 2018]{talvard2018} Talvard, M., Rev{\'e}ret, V., Le-Pennec, Y., et al.\ 2018, \procspie, 10708, 1070838

\bibitem[Teixeira et al. 2016]{teixeira2016} Teixeira, P.~S., Takahashi, S., Zapata, L.~A., et al.\ 2016, \aap, 587, A47

\bibitem[Urquhart et al. 2022]{urquhart2022} Urquhart, J.~S., Wells, M.~R.~A., Pillai, T., et al.\ 2022, \mnras, 510, 3389

\bibitem[Walker et al. 2021]{walker2021} Walker, D.~L., Longmore, S.~N., Bally, J., et al.\ 2021, \mnras, 503, 77

\bibitem[Wang et al. 2014]{wang2014} Wang, K., Zhang, Q., Testi, L., et al.\ 2014, \mnras, 439, 3275

\bibitem[Wenger et al. 2018]{wenger2018} Wenger, T.~V., Balser, D.~S., Anderson, L.~D., et al.\ 2018, \apj, 856, 52

\bibitem[Werner et al. 2004]{werner2004} Werner, M.~W., Roellig, T.~L., Low, F.~J., et al.\ 2004, \apjs, 154, 1.

\bibitem[Wilcock et al. 2012]{wilcock2012} Wilcock, L.~A., Ward-Thompson, D., Kirk, J.~M., et al.\ 2012, \mnras, 422, 1071

\bibitem[Williams et al. 1994]{williams1994} Williams, J.~P., de Geus, E.~J., \& Blitz, L.\ 1994, \apj, 428, 693

\bibitem[Yu et al. 2018]{yu2018} Yu, N.-P., Xu, J.-L., \& Wang, J.-J.\ 2018, Research in Astronomy and Astrophysics, 18, 015

\bibitem[Yuan et al. 2020]{yuan2020} Yuan, L., Li, G.-X., Zhu, M., et al.\ 2020, \aap, 637, A67




\end{thebibliography}
\end{document}